\documentclass[12pt]{article}
\usepackage{latexsym}
\usepackage{epsfig,amssymb,euscript}
\usepackage{amsmath}
\usepackage{array,calc,epsfig}
\usepackage{citesort}
\usepackage{xypic}

\def\be{\begin{equation}}
\def\bea{\begin{eqnarray}}
\def\eea{\end{eqnarray}}

\numberwithin{equation}{section} 
\usepackage[]{graphicx}

\def\calb         {{\cal B}}
\def\calc         {{\cal C}}
\def\cald         {{\cal D}}

\def\calf         {{\cal F}}
\def\calg         {{\cal G}}

\def\cali         {{\cal I}}
\def\calj         {{\cal J}}
\def\calk         {{\cal K}}
\def\call         {{\cal L}}
\def\calm         {{\cal M}}
\def\caln         {{\cal N}}

\def\calt         {{\cal T}}

\def\calv         {{\cal V}}
\def\calw         {{\cal W}}

\def\Im           {{\rm Im\hskip0.1em}}

\def\sqr#1#2{{\vcenter{\vbox{\hrule height.#2pt
 \hbox{\vrule width.#2pt height#1pt \kern#1pt \vrule width.#2pt}\hrule
 height.#2pt}}}}

\def\slashchar#1{\setbox0=\hbox{$#1$}           
\dimen0=\wd0                                 
\setbox1=\hbox{/} \dimen1=\wd1               
\ifdim\dimen0>\dimen1                        
\rlap{\hbox to \dimen0{\hfil/\hfil}}      
#1                                        
\else                                        
\rlap{\hbox to \dimen1{\hfil$#1$\hfil}}   
/                                         
\fi}

\begin{document}
\font\cmss=cmss10 \font\cmsss=cmss10 at 7pt
\leftline{\tt hep-th/0610044}

\vskip -0.5cm
\rightline{\small{\tt MPP-2006-126}}
\rightline{\small{\tt KUL-TF-06/25}}

\vskip .7 cm

\hfill
\vspace{18pt}
\begin{center}
{\Large \textbf{Deformations of calibrated D-branes }}

\smallskip

{\Large \textbf{in flux generalized complex manifolds}}
\end{center}

\vspace{6pt}
\begin{center}
{\large\textsl{Paul Koerber~$^{a}$ and Luca Martucci~$^{b}$}}

\vspace{25pt}
\textit{\small $^a$ Max-Planck-Institut f\"ur Physik\\
F\"ohringer Ring 6, D-80805 Munich, Germany}\\
\vspace{6pt}
\textit{\small $^b$ Institute for Theoretical Physics, K.U. Leuven,\\
Celestijnenlaan 200D, B-3001 Leuven, Belgium}
\end{center}

\vspace{12pt}

\begin{center}
\textbf{Abstract}
\end{center}

\vspace{4pt} {\small \noindent

We study massless deformations of generalized calibrated cycles, which describe, in the language of generalized
complex geometry, supersymmetric D-branes in $\caln=1$ supersymmetric compactifications with fluxes.
We find that the deformations are classified by the first cohomology group of a Lie algebroid
canonically associated to the generalized calibrated cycle, seen as a generalized complex submanifold with respect to the
integrable generalized complex structure of the bulk. We provide examples in the $SU(3)$ structure case and
in a `genuine' generalized complex structure case. We discuss cases of  lifting of massless modes  due to world-volume fluxes,
background fluxes and a generalized complex structure that changes type.

}

\vspace{4cm}


\vfill
\vskip 5.mm
\hrule width 5.cm
\vskip 2.mm
{\small \noindent e-mails: koerber@mppmu.mpg.de, luca.martucci@fys.kuleuven.be}

\newpage

\tableofcontents

\section{Introduction}

The study of string theory compactifications with fluxes has gained a considerable amount of momentum in recent years due to the
central place they seem to take in the search for phenomenologically interesting models.
More in particular, since they allow for some or even all of the moduli to be stabilized they play a central role as the building blocks of a
discrete string theory landscape. The current paradigm is to aim for models that preserve $\caln=1$ supersymmetry in the four flat directions
that make up our world. The reasons for this are that supersymmetric models are under much better control, while also from particle physics
a model with $\caln=1$ matter sector in which the supersymmetry is broken at lower energy seems to be preferred.

Much of the efforts so far have been focused on the special case of $SU(3)$ structure vacua, of which the most studied subcase is
warped Calabi-Yau in type IIB string theory (for a review and more references see \cite{granareview}).
It was realized in \cite{gmpt} (see \cite{fidanza,granaold,witt0} for earlier work) that the appropriate tool for studying the most general vacua preserving
four-dimensional Poincar\'e invariance and $\caln=1$ supersymmetry is generalized complex geometry, of which the mathematical foundations were laid shortly before in \cite{hitchin,gualtieri}.
The internal compactified manifold $M$ has then instead of reduced $SU(3)$ structure on its tangent bundle $T_M$ reduced $SU(3) \times SU(3)$
structure on $T_M \oplus T^\star_M$ and is characterized by two pure spinors $\hat{\Psi}_1$ and $\hat{\Psi}_2$. One of them, $\hat{\Psi}_2$
corresponds to an {\em integrable} generalized complex structure. The other pure spinor $\hat{\Psi}_1$ corresponds to a generalized almost
complex structure which is in general not integrable in the presence of R-R fluxes.

It is hard to construct concrete examples of flux vacua with compact internal space --- basically the reason is a no-go theorem \cite{nogo} which requires the introduction of sources with negative tension
(like orientifold planes) --- and it is even more so to find genuine $SU(3) \times SU(3)$ structure examples that are not already in the
$SU(3)$ structure case. Orientifolds in the generalized setting were introduced in \cite{grimm} and a systematic search for genuine $SU(3) \times SU(3)$
structure manifolds based on nilmanifolds and solvmanifolds was performed in \cite{grananil}.

So far for the bulk story. A fundamental role in type II string theories is however played by D-branes.
For instance, one needs D-branes to provide the matter sector in phenomenological models. In this paper we want to study the moduli of BPS D-branes, i.e. the D-branes that preserve the $\caln=1$ supersymmetry
of the background. It turns out that, in addition to generalized complex geometry being the appropriate tool to study the supersymmetry conditions of the
bulk, it is equally applicable to the conditions for D-branes in such a background to be supersymmetric. Indeed, in \cite{koerber}
it was shown, in backgrounds with only NSNS flux and $\caln=2$ supersymmetry, that for a D-brane to preserve supersymmetry it has to be
generalized calibrated and this was extended in \cite{lucal} to backgrounds with RR flux and $\caln=1$ supersymmetry. These works generalized
the earlier concept of calibrations in backgrounds with fluxes, also dubbed generalized calibrations\footnote{Hoping that this does not cause confusion,
when we talk about generalized calibrations in this paper, we mean the calibrations of \cite{koerber,lucal}, naturally embedded in generalized complex
geometry.} and introduced in the seminal papers \cite{gencalold}, to include the world-volume gauge field and naturally embedded it in the language of
generalized complex geometry.
In \cite{lucal} it was also shown that supersymmetric D-branes are only allowed in backgrounds where the norms of the two  internal spinors associated to the $SU(3)\times SU(3)$ structure are equal. Backgrounds satisfying this extra condition were dubbed {\em D-calibrated} in \cite{branesuppot}.

Take a D-brane wrapping a cycle $\Sigma$ with a field strength $\calf$ on it,
then we say in the language of generalized complex geometry that the D-brane wraps the generalized cycle $(\Sigma,\calf)$.
It is generalized calibrated with respect to ${\rm Re} \, \hat{\Psi}_1$ if it wraps
a generalized complex cycle with respect to the integrable generalized complex structure corresponding to $\hat{\Psi}_2$ and additionally the top form in $ P_\Sigma[{\rm \Im}\, \hat{\Psi}_1]\wedge e^\calf$
vanishes on the cycle. So we see that the generalized calibration condition decouples in a pair of
conditions, which in fact have a clear 4-dimensional interpretation as an F-flatness and a D-flatness condition \cite{branesuppot}.
If both conditions are satisfied, up to an appropriate choice of orientation the D-brane saturates a calibration bound
in that its Dirac-Born-Infeld action reduces to an integral of a form proportional to the top form in $P_\Sigma[{\rm Re} \, \hat{\Psi}_1] \wedge e^\calf$.
In an $SU(3)$ structure background this general definition reduces to something very similar to the well-known conditions for supersymmetric D-branes on a standard Calabi-Yau.
In particular, as in a standard Calabi-Yau, the F-flatness condition restricts for type IIB to B-branes, complex D-branes with $\calf_{0,2}=0$,
and in type IIA to A-branes, Lagrangian cycles with $\calf=0$ or more general coisotropic branes \`a la \cite{kapustincoisotropic,kapustinstability}.
On the other hand the background fluxes affect the D-flatness/stability condition through $\hat{\Psi}_1$.

In this paper we study the conditions for infinitesimal deformations to preserve the generalized calibration condition\footnote{The deformation theory of  calibrations of the kind considered in \cite{gencalold} is studied in \cite{defold}.}. These deformations will
thus transform supersymmetric D-branes into supersymmetric D-branes and correspond to massless fluctuations. Our main result is that
the massless deformations are counted by the Lie algebroid cohomology group $H^1(L_{(\Sigma,\calf)})$ where $L_{(\Sigma,\calf)}$ is the intersection
of $T_{(\Sigma,\calf)}$, the generalized tangent bundle of the D-brane, with $L|_\Sigma$, the $+i$-eigenbundle of $\calj$, the integrable
generalized complex structure corresponding to the pure spinor $\hat{\Psi}_2$. So we immediately note that this cohomology group depends
on only {\em one} of the pure spinors describing the background geometry, so in a sense on only half of the bulk data,
even if we have seen above that the second part of the calibration condition {\em does} depend on $\hat{\Psi}_1$.
The reason is clear rephrasing the problem in an $\caln=1$ four-dimensional description,
where as usual the superpotential, here depending only on $\hat{\Psi}_2$, is the only information one needs to describe the moduli space of a theory.
In the present case, this expected fact translates as follows.
We will find that the deformations that preserve generalized complex cycles, the first part of the calibration condition, should satisfy the
condition that they are described by closed 1-forms on $L_{(\Sigma,\calf)}$. But we will also observe that this condition has an extended gauge symmetry, which we obtain by using $\calj$ to complexify the real gauge symmetry on the world-volume. This extended gauge symmetry is generated by exact 1-forms on $L_{(\Sigma,\calf)}$.
Now, the second part of the calibration condition provides for a gauge-fixing of the imaginary part of the extended gauge symmetry.
So we find that there is one and only one massless deformation in each equivalence class $H^1_{(\Sigma,\calf)}$,
which is explicitly identified by the complexification of the second part of the calibration condition and thus $\hat{\Psi}_1$. In fact, we will find it is given by the harmonic representative of $H^1(L_{(\Sigma,\calf)})$.

From a mathematical point of view we are essentially addressing the problem of generalizing
the known results on deformations of complex \cite{kodaira} and special Lagrangian cycles \cite{mclean},
focusing in this paper on the first order infinitesimal deformations (the massless modes from a four-dimensional point of view). For example, in  \cite{mclean}   it was found that the calibration preserving deformations of a special Lagrangian cycle are characterized by $H^1(M,\mathbb{R})$. Moreover, they are in fact given by the harmonic representatives.
We will rediscuss this as a special case of our general formula reaching naturally the well-known result that  $H^1(M,\mathbb{R})$ must be complexified to $H^1(M,\mathbb{C})$ to include the Wilson line moduli, consistent with supersymmetry.

As we mentioned above, we start by studying deformations of generalized complex cycles, which are also important from
the point of view of topological string theory with fluxes first introduced in \cite{topo1} and expanded upon in \cite{topo2,topo3,topo4}.
Indeed, generalized complex D-branes are precisely the ones that preserve the BRST operator and are thus the consistent ones in the topological string theory \cite{kapustinstability,zabzineDbranes}.
In \cite{kapustindeform} the BRST cohomology of open strings with boundary conditions given by such a D-brane was calculated and found to be exactly
given by the cohomology groups $H^k(L_{(\Sigma,\calf)})$. The observables that correspond to deformations of the boundary are precisely
those for which $k=1$. So we find that Kapustin's result agrees with ours if D-branes related by a complexified gauge
transformation in our formalism are also equivalent as boundary conditions in the topological string theory. In the special Lagrangian case this
is the statement that A-branes differing by a Hamiltonian deformation are equivalent.

We will study examples in the Calabi-Yau case, the $SU(3)$ structure case and also in a background defined by a genuine generalized complex structure.
We will (re)discuss in our setting mechanisms for moduli-lifting due to world-volume fluxes as well as background fluxes. But interestingly, some moduli
can also be lifted merely by the choice of a non-trivial background generalized complex structure that changes type. We will show
this for a D3-brane. Although point-like in the internal manifold, we will still find non-trivial cohomology.

In section \ref{review} we review background material on generalized complex geometry and generalized submanifolds. For a more
extensive review that is very readable also for physicists we refer to \cite{gualtieri}. In section \ref{deformations} we study the condition
for infinitesimal deformations to preserve generalized complex cycles, while in section \ref{gauge} we introduce the complexified gauge symmetry
and derive the result that the deformations are classified by the cohomology group $H^1(L_{(\Sigma,\calf)})$. In section \ref{Dterm} we study deformations of the second
part of the generalized calibration condition and arrive at a gauge-fixing of the complexified gauge symmetry. We comment in section \ref{4d} on
the more difficult issue of higher order deformations and establish the link with the superpotential of \cite{branesuppot}. In section \ref{examples} we discuss examples. We build up from the well-studied fluxless Calabi-Yau case, to the Calabi-Yau
case with world-volume fluxes to arrive at the $SU(3)$ structure case with background fluxes. We conclude with an example of
a D3-brane in an honest $SU(3) \times SU(3)$ background. We defer some more technical issues to the appendixes. In appendix
\ref{mm} we provide the proofs for the statements on gauge-fixing in section \ref{Dterm} and we define the appropriate metric on Lie algebroid forms and discuss the derived codifferential operator in appendix \ref{hodge}. In appendix \ref{masses} we calculate the classical masses and show that the massless deformations are indeed the calibration/supersymmetry preserving ones. We comment on stability in appendix \ref{stability}. Finally, in appendix \ref{k} we discuss the general formula for the K\"ahler potential for small fluctuations of a space-time filling D-brane around a supersymmetric configuration.

\section{Preliminary remarks on generalized submanifolds and generalized complex geometry}
\label{review}


This section has the aim to introduce some background material necessary for the study of
the deformations of generalized complex and generalized calibrated cycles, which will be introduced in the next section.

In  subsection \ref{sub1} we recall the definition of a generalized submanifold given in \cite{gualtieri},
and introduce moreover the notion of a generalized current, which will allow to simplify the analysis of the rest of the paper.
In subsection \ref{sub2} we briefly review some basic concepts of generalized complex geometry and the Hodge-like decomposition of forms \cite{hitchin,gualtieri,cavalcanti}. We will keep the analysis general, considering ambient manifolds of arbitrary even dimension $d$.

\subsection{Generalized submanifolds and generalized currents}
\label{sub1}

The key point about generalized geometry is to consider the vector bundle $T_M \oplus T^\star_M$ instead of $T_M$. So take $T_M\oplus T^\star_M$ on a  even $d$-dimensional manifold $M$ (with coordinates $y^m,\ m=1,\ldots,d$).
 On each fiber of $T_M\oplus T^\star_M$ there is a canonical metric $\cali$ of signature $(d,d)$ defined in the following way:
 given vectors $\mathbb{X}=X+\xi$, $\mathbb{Y}=Y+\eta$ in a fiber of $T_M\oplus T^\star_M$ we have that
\bea
\cali(\mathbb{X},\mathbb{Y})=\frac{1}{2}(\eta(X)+\xi(Y)) \ .
\eea

In our discussion, we will always consider the general possibility of having  non-trivial NSNS closed 3-form $H$
(which in string theory must obey a proper quantization condition) on $M$. Then, as defined in \cite{gualtieri}, a generalized submanifold consists of a pair $(\Sigma,\calf)$ of a submanifold $\Sigma \subset M$ and a 2-form $\calf$ on $\Sigma$ such that $d\calf = P_\Sigma[H]$. The generalized tangent bundle $T_{(\Sigma,\calf)}$ of a generalized submanifold is defined as follows
\begin{equation}\label{gtb}
T_{(\Sigma,\calf)} = \{X+ \xi \in T_\Sigma \oplus T^\star_M|_\Sigma : P_\Sigma[\xi] = \iota_X \calf \}\ .
\end{equation}
$T_{(\Sigma,\calf)}$ is a real maximal isotropic sub-bundle of the restricted bundle $T_M\oplus T^\star_M|_\Sigma$
\footnote{Isotropic means that $\cali(\mathbb{X},\mathbb{Y})=0$ for any $\mathbb{X},\mathbb{Y}\in T_{(\Sigma,\calf)}|_{p\in \Sigma}$, maximal
means it has the maximal dimension $d$.}.

It is now convenient to introduce the notion of generalized currents,
defined as linear maps on the space of differentiable polyforms on $M$ \footnote{If $M$ is non-compact
one can require the smooth polyforms to have suitable asymptotic behaviour.}.
A generalized {\em real} current $j$ can be formally seen as a polyform
(which we indicate with the same symbol $j$)  such that for any smooth polyform $\phi$ we have
\bea
j(\phi)\equiv\int_M\langle\phi,j\rangle \ ,
\eea
with $\langle\cdot,\cdot\rangle$ denoting the Mukai pairing defined as
\begin{equation}
\langle \phi_1 , \phi_2 \rangle = \phi_1 \wedge \sigma(\phi_2)|_{\text{top}} \ ,
\end{equation}
where $\sigma$ reverses the indices of a $k$-form $\phi=\frac{1}{k!}\phi_{m_1\ldots m_k}dy^{m_1}\wedge\ldots\wedge dy^{m_k}$:
\bea
\sigma(\phi)=\frac{1}{k!}\phi_{m_1\ldots m_k}dy^{m_k}\wedge\ldots\wedge dy^{m_1}\ .
\eea
We will only consider polyforms of definite parity, i.e.\ the forms of different dimensions making up the polyform are of either
even or odd dimension. And likewise for the currents.

The Mukai pairing satisfies the two following immediate properties. First, for any pair of smooth polyforms $\phi_1,\phi_2$ of definite opposite parity, we have
\bea\label{prop1}
\int_M \langle d_H\phi_1 , \phi_2 \rangle =\int_M \langle \phi_1 , d_H\phi_2 \rangle\ ,
\eea
where $d_H=d + H \wedge$ is the $H$-twisted exterior derivative.
 Secondly, recalling that the Clifford action of a generalized vector $\mathbb{X}=X+\xi\in T_M\oplus T^\star_M$ on a polyform $\phi$ is given by $\mathbb{X}\cdot \phi=\iota_X\phi+\xi\wedge\phi$, we have that
\bea\label{prop2}
\langle \mathbb{X}\cdot\phi_1 , \phi_2 \rangle=-\langle \phi_1 , \mathbb{X}\cdot\phi_2 \rangle\ .
\eea
The properties (\ref{prop1}) and (\ref{prop2}) allow to extend the action of the twisted exterior derivative $d_H$ and of a generalized vector $\mathbb{X}$
to the space of generalized currents in the same way as for standard currents (see for example \cite{GH}).

We can associate a current $j_{(\Sigma,\calf)}$ (of definite parity) to a generalized submanifold $(\Sigma,\calf)$, acting on a general polyform $\phi$ (of the same parity) in the following way\footnote{Note that $j_{(\Sigma,\calf)}$ has support on $\Sigma$, where by support of a current $j$, denoted supp$(j)$, one means the smallest closed set in $M$ such that $j(\phi)=0$ for any smooth polyform $\phi$ with compact support on $M-{\rm supp}(j)$.}
\bea
\int_M\langle \phi,j_{(\Sigma,\calf)}\rangle \equiv \int_\Sigma P_{\Sigma}[\phi]\wedge e^\calf\ .
\eea

Since for any smooth polyform $\phi$ we have that
\bea
\int_M\langle \phi,d_Hj_{(\Sigma,\calf)}\rangle\equiv \int_M\langle d_H\phi,j_{(\Sigma,\calf)}\rangle=\int_\Sigma P_{\Sigma}[d_H\phi]\wedge e^\calf=\int_{\partial\Sigma} P_{\partial\Sigma}[\phi]\wedge e^{P_{\partial\Sigma}[\calf]}\ ,
\eea
we see that $d_Hj_{(\Sigma,\calf)}=j_{(\partial\Sigma,P_{\partial\Sigma}[\calf])}$ and, in particular, if $\Sigma$ is a cycle then
\bea
d_Hj_{(\Sigma,\calf)}=0\ .
\label{currentclosed}
\eea

Furthermore, we say that two generalized cycles $(\Sigma,\calf)$ and $(\Sigma^\prime,\calf^\prime)$
are in the same generalized homology class if there exists a generalized submanifold $(\tilde\Sigma,\tilde\calf)$ such that $\partial{\tilde\Sigma}=\Sigma^\prime-\Sigma$ with $P_\Sigma[\tilde\calf]=\calf$ and $P_{\Sigma^\prime}[\tilde\calf]=\calf^\prime $. It is easy to see that in this case
\bea
j_{(\Sigma^\prime,\calf^\prime)}-j_{(\Sigma,\calf)}=d_Hj_{({\tilde\Sigma},\tilde\calf)}\ .
\eea
Thus if $(\Sigma,\calf)\sim(\Sigma^\prime,\calf^\prime)$ in generalized homology, then $j_{(\Sigma^\prime,\calf^\prime)}\sim j_{(\Sigma,\calf)}$ in $d_H$-cohomology and a generalized homology class $[(\Sigma,\calf)]$ determines a $d_H$-cohomology class $[j_{(\Sigma,\calf)}]\in H^\bullet_{d_H}(M)$.\footnote{\label{note1}Note that, even if  we have in fact used currents, we obtain ordinary $d_H$-cohomology elements due to the isomorphism between de Rham cohomology and current cohomology \cite{GH} which we expect to still hold in the $H$-twisted case.}

The generalized current associated to a generalized cycle allows us to give the following alternative characterization of the generalized tangent bundle.
We have already said that the generalized tangent bundle $T_{(\Sigma,\calf)}$ to a generalized cycle $(\Sigma,\calf)$ defined in (\ref{gtb})
is a maximal isotropic sub-bundle of  $T_M\oplus T^\star_M|_\Sigma$.
This can alternatively be defined as the sub-bundle   of  $T_M\oplus T^\star_M|_\Sigma$ whose sections $\mathbb{X}$ annihilate $j_{(\Sigma,\calf)}$.
To properly define this one first extends $\mathbb{X}$ to a section of $T_M\oplus T^\star_M$,
and then defines $\mathbb{X}\cdot j_{(\Sigma,\calf)}$ as follows. For any smooth polyform $\phi$ of appropriate definite parity we have
\bea
\int_M\langle \phi,\mathbb{X}\cdot j_{(\Sigma,\calf)}\rangle\equiv -\int_M\langle \mathbb{X}\cdot\phi,j_{(\Sigma,\calf)}\rangle\ .
\eea
Since $j_{(\Sigma,\calf)}$ has support $\Sigma$, the definition of $\mathbb{X}\cdot j_{(\Sigma,\calf)}$ does not actually depend on the extension of $\mathbb{X}$ outside $\Sigma$.
To show that this definition indeed coincides with the earlier one we observe
that if we take a section $\mathbb{X}=X+\xi$ of $T_{(\Sigma,\calf)}$ we find for any polyform $\phi$
\begin{equation}
\int_M\langle \mathbb{X}\cdot\phi,j_{(\Sigma,\calf)}\rangle\ = \int_\Sigma \iota_X \left(P_\Sigma[\phi] \wedge e^\calf\right) = 0 \, .
\end{equation}
There cannot be any other $\mathbb X$ annihilating $j_{(\Sigma,\calf)}$ because $T_{(\Sigma,\calf)}$ already has the maximal dimension
for a space of annihilators of a spinor. Roughly for a $k$-cycle the generalized current $j_{(\Sigma,\calf)}$ looks like $\delta^{(d-k)}(\Sigma)\wedge e^{-\calf}$, where $\delta^{(d-k)}(\Sigma)$ is the ordinary current associated to the cycle $\Sigma$.
In summary, $j_{(\Sigma,\calf)}$ can be thought of as a sort of localized pure spinor associated to $T_{(\Sigma,\calf)}$, seen as a maximal isotropic sub-bundle of $T_M\oplus T^\star_M|_\Sigma$.

\subsection{Generalized complex manifolds and the decomposition of forms}
\label{sub2}

A generalized almost complex structure on $M$ is given by a fiberwise map
\bea
\calj: T_M\oplus T^\star_M\rightarrow T_M\oplus T^\star_M \ ,
\eea
such that
\bea
\calj^2=-1\quad {\rm and} \quad \cali(\calj\mathbb{X},\calj\mathbb{Y})=\cali(\mathbb{X},\mathbb{Y}) \, .
\eea
$\calj$  defines a(n integrable) generalized complex structure if its $+i$-eigenbundle $L\subset (T_M\oplus T^\star_M)\otimes \mathbb{C}$ is involutive with respect to the $H$-twisted Courant bracket defined as follows on sections $X+\xi,Y+\eta$ of $T_M\oplus T^\star_M$ \cite{hitchin,gualtieri}
\begin{equation}\label{Hcourant}
[X+\xi,Y+\eta]_H=[X,Y]+ \call_X \eta - \call_Y \xi - \frac{1}{2} d(\iota_X \eta - \iota_Y \xi) + \iota_X \iota_Y H \ ,
\end{equation}
where $[\cdot,\cdot]$ is the standard Lie bracket on sections of $T_M$.
For further use we recall the following property of the Courant bracket
\begin{equation}
[X+\iota_X B, Y+\iota_Y B]_{dB} = [X,Y]+\iota_{[X,Y]}B \, .
\label{propcourant}
\end{equation}
In \cite{gualtieri} a generalized Darboux theorem was proven stating that a generalized complex structure implies the existence of local hybrid complex-symplectic coordinates.

Any generalized almost complex structure is associated to a locally defined pure spinor $\psi$ which is annihilated by $L$ (see \cite{gualtieri} for more details).
If there is a globally defined pure spinor satisfying $d_H\psi=0$ (one can show that from this integrability follows \cite{hitchin}) then $M$ is,
using Hitchin's terminology, a (weak) generalized Calabi-Yau manifold.

In a generalized almost complex manifold $M$ we have the following Hodge-like decomposition of the space of forms (see \cite{gualtieri,gualtieri2,cavalcanti}):
\bea\label{dec}
\Lambda^\bullet T^\star_M\otimes\mathbb{C}=\bigoplus_{k=-d/2}^{d/2} U_k\ ,
\eea
where
\bea
U_k=\Lambda^{d/2-k}\bar L\cdot \psi\ .
\eea
Using the natural metric $\cali$ on $T_M\oplus T^\star_M$, one can think of $\calj$ as a section of $so(T_M\oplus T_M^\star)$ that acts via the spin representation on the polyforms on $M$.
Thus, one can give an alternative definition of $U_k$ as the $ik$-eigenbundle of $\calj$. Note that $\psi\in\Gamma(U_{d/2})$ and
for $\phi \in \Gamma(U_k)$ we have $\bar\phi\in\Gamma(U_{-k})$.  Moreover if $\phi_1\in \Gamma(U_k)$, then
\bea
\langle \phi_1,\phi_2\rangle=0\quad {\rm if} \quad \phi_2|_{U_{-k}}=0\ ,
\label{mukcomp}
\eea
 with $\phi_2|_{U_{-k}}\equiv \pi_{-k}(\phi_2)$, where
\bea
\pi_k:\Lambda^\bullet T^\star_M\otimes \mathbb{C}\rightarrow U_k\ ,
\eea
 denotes the projector on $U_k$.

One can also define $\partial_H$ and $\bar\partial_H$ as given by $\pi_{k+1}\circ d_H$ and $\pi_{k-1}\circ d_H$ respectively on $\Gamma(U_k)$.
It is possible to see \cite{gualtieri} that the ($H$-twisted) integrability of $\calj$ is equivalent to the requirement that
\bea
d_H=\partial_H+\bar\partial_H\ .
\eea

With property \eqref{mukcomp} in mind the decomposition (\ref{dec}) can be extended to the generalized currents in the following way.
One defines a current $j$ as a ``distributional'' section of $U_k$  (in symbols, $j\in\Gamma_{\rm cur}(U_k)$) if $j(\phi)=0$ for any (smooth) polyform $\phi\in \Gamma(U_{p\neq -k})$.

\section{Generalized complex submanifolds and their first order deformations}
\label{deformations}

Let $M$ be a generalized complex manifold.
Then a generalized complex submanifold \cite{gualtieri} is given by a generalized submanifold $(\Sigma,\calf)$
such that its generalized tangent bundle $T_{(\Sigma,\calf)}$ is stable under $\calj$.
We can give an alternative equivalent form for this condition, as follows:
a generalized submanifold $(\Sigma,\calf)$ is generalized complex
if and only if $j_{(\Sigma,\calf)}$ is a distributional section of $U_0$, i.e.
\bea\label{currcond}
(\Sigma,\calf)\ {\rm generalized\ complex\ submanifold}\quad\Leftrightarrow\quad j_{(\Sigma,\calf)}\in \Gamma_{\rm cur}(U_0)\ .
\eea
Indeed, in \cite{cavalcanti} it was proven that in any point the pure spinor of a $\calj$-invariant real maximal isotropic subspace is an element of $U_0$.

\smallskip

Let us now consider the possible infinitesimal deformations of generalized complex cycles.
As discussed in \cite{branesuppot} a general infinitesimal deformation of a generalized cycle $(\Sigma,\calf)$ is described by sections of the {\em generalized normal bundle} defined as $\caln_{(\Sigma,\calf)}\equiv (T_M\oplus T^\star_M)|_\Sigma/T_{(\Sigma,\calf)}$.
We indicate the sections of $\caln_{(\Sigma,\calf)}$ with $[\mathbb{X}]$ or equivalently, when there is no confusion nor inconsistencies, with representative sections $\mathbb{X}$ of $T_M\oplus T^\star_M|_\Sigma$.
If we restrict to {\em generalized complex} cycles there is, since in this case $T_{(\Sigma,\calf)}$ is stable under $\calj$, a canonical fiberwise defined complex structure on $\caln_{(\Sigma,\calf)}$
induced by $\calj$ \cite{branesuppot}.  We can thus
introduce the holomorphic and anti-holomorphic generalized normal bundles $\caln_{(\Sigma,\calf)}^{1,0}$ and $\caln_{(\Sigma,\calf)}^{0,1}$ and
split a generalized normal vector  $[\mathbb{X}]$ in holomorphic and antiholomorphic parts: $[\mathbb{X}]=[\mathbb{X}]^{1,0}+[\mathbb{X}]^{0,1}$.

In general, a section $[\mathbb{X}]$ of the generalized normal bundle to a generalized cycle $(\Sigma,\calf)$ defines,
via a representative generalized vector field $\mathbb{X}=X+\xi$,
a deformation of the cycle $(\Sigma,\calf)$ consisting of the infinitesimal diffeomorphism defined by $X$
(which acts also on $\calf$ as $\delta_X\calf=P_\Sigma[\iota_X H]$, see \cite{branesuppot}), and a deformation of the field-strength $\calf$ defined by $\delta_\xi\calf=dP_{\Sigma}[\xi]$.
In this way one can easily see how the quotient with respect to $T_{(\Sigma,\calf)}$ in the definition of $\caln_{(\Sigma,\calf)}$ corresponds
to considering deformations related by world-volume diffeomorphisms as equivalent.

Now, we would like to determine under which conditions a section $[\mathbb{X}]\in\Gamma[\caln_{(\Sigma,\calf)}]$ generates an infinitesimal deformation of a generalized complex cycle $(\Sigma,\calf)$
into another generalized complex cycle (at first order).
To do this, we will translate the problem in terms of the dual generalized current $j_{(\Sigma,\calf)}$.
The infinitesimal deformation $\delta_{\mathbb{X}}j_{(\Sigma,\calf)}$  of the current induced by the generalized normal vector field $[\mathbb{X}]=[X+\xi]$
is defined by its action on a general smooth polyform $\phi$
\bea
\delta_{[\mathbb{X}]}j_{(\Sigma,\calf)}(\phi)&\equiv&\int_{M}\langle \phi,\delta_{[\mathbb{X}]}j_{(\Sigma,\calf)}\rangle \, ,
\eea
and
\bea
\label{varj}
\delta_{[\mathbb{X}]}j_{(\Sigma,\calf)}(\phi)&\equiv& \int_{\Sigma + \delta \Sigma} P_{\Sigma+\delta \Sigma}[\phi] \wedge e^\calf + \int_\Sigma P_\Sigma[\phi] \wedge \delta \calf \wedge e^\calf - \int_\Sigma P_\Sigma[\phi] \wedge e^\calf \cr
&=&\int_\Sigma\big\{ P_\Sigma[\call_X\phi+\phi\wedge\iota_XH + \phi\wedge d\xi ]\big\}\wedge e^\calf\cr
&=& \int_\Sigma P_\Sigma[\mathbb{X} \cdot d_H\phi + d_H (\mathbb{X} \cdot \phi)]\wedge e^\calf =  -\int_M\langle \phi, d_H(\mathbb{X}\cdot j_{(\Sigma,\calf)})\rangle\ .
\eea
Here we used that $\delta \calf = \delta_X \calf + \delta_\xi \calf = P_\Sigma[\iota_X H] + dP_\Sigma[\xi]$,
the properties \eqref{prop1} and \eqref{prop2} of the Mukai pairing and \eqref{currentclosed}.
Note that all previous expressions are invariant under the shift $\mathbb{X}\rightarrow \mathbb{X}+\mathbb{Y}$ with $\mathbb{Y}$ a section of $T_{(\Sigma,\calf)}$
so that they don't depend on the choice of representative of $[\mathbb{X}]$.

We see it is natural to define
\begin{equation}
\call_{\mathbb{X}} \phi \equiv \mathbb{X} \cdot d_H\phi + d_H (\mathbb{X} \cdot \phi) \, ,
\end{equation}
and we obtain for the deformation of the current under $[\mathbb{X}]$
\bea
\delta_{[\mathbb{X}]}j_{(\Sigma,\calf)}\equiv - \call_{\mathbb{X}} j_{(\Sigma,\calf)} = -d_H(\mathbb{X}\cdot j_{(\Sigma,\calf)}) ,
\eea
where again this expression must be considered in the distributional sense in which case it is well defined for any representative $\mathbb{X}$ of $[\mathbb{X}]$.

Imposing now that the infinitesimally deformed cycle is still generalized complex we find, using the characterization of generalized complex in (\ref{currcond}),
the following concise condition
\bea\label{defcond}
\bar\partial_H(\mathbb{X}^{0,1}\cdot j_{(\Sigma,\calf)})=0\ ,
\eea
or equivalently its complex conjugate $\partial_H(\mathbb{X}^{1,0}\cdot j_{(\Sigma,\calf)})=0$.
As we will show more explicitly in section \ref{examples},
the condition (\ref{defcond}) is the direct generalization of the standard result that first order deformations of a complex submanifold are given by holomorphic sections of the ordinary holomorphic normal bundle.
For this reason we will denote the space of sections of $\caln^{0,1}_{(\Sigma,\calf)}$ satisfying the condition (\ref{defcond})
by $\Gamma_{\rm hol}(\caln_{(\Sigma,\calf)}^{0,1})$.

\section{From gauge equivalent deformations to Lie algebroid cohomology}
\label{gauge}

Until now we have considered all the possible deformations of a generalized complex cycle,
already taking into account the equivalence under world-volume diffeomorphisms. However, in the study of the moduli space of generalized calibrations,
one must also consider another equivalence relation between different generalized cycles \cite{branesuppot}, which naturally preserves the generalized holomorphicity of the generalized complex cycles. This can be introduced as follows. We have already described the possible infinitesimal deformations as global sections of $\caln_{(\Sigma,\calf)}$.
However, there is a subclass of transformations that clearly leave the generalized cycles unchanged.
In D-brane language, these are given by infinitesimal gauge transformations of the world-volume gauge field.
If $\lambda$ denotes the gauge parameter, such transformations are generated by a generalized normal vector field $[\mathbb{X}_\lambda]$
with representative $\mathbb{X}_\lambda=\xi_\lambda$, with $\xi_\lambda\in \Gamma[T^\star_M|_\Sigma]$ such that $P_\Sigma[\xi_\lambda]=d\lambda$.  Now, such a generalized normal vector field obviously satisfies the condition (\ref{defcond}). Thus, it is evident that also the generalized normal vector field $[\calj\mathbb{X}_\lambda]$ satisfies the condition (\ref{defcond}) and then defines a deformation which preserves the generalized holomorphicity condition of the generalized cycles.
$\calj[\mathbb{X}_\lambda]$ can be considered as the `imaginary' extension of the real infinitesimal gauge transformation generated by $[\mathbb{X}_\lambda]$.
$[\mathbb{X}_\lambda]$ and $\calj[\mathbb{X}_\lambda]$ together generate the space of complexified infinitesimal gauge transformations ${\bf g}^{\mathbb{C}}_{(\Sigma,\calf)}\subset \Gamma(\caln_{(\Sigma,\calf)})$.
We will consider infinitesimal deformations related by an element in ${\bf g}^{\mathbb{C}}_{(\Sigma,\calf)}$ as equivalent,
and the physically distinguishable first order deformations are then given by $\Gamma_{\rm hol}(\caln_{(\Sigma,\calf)}^{0,1})/({\bf g}^{\mathbb{C}}_{(\Sigma,\calf)})^{0,1}$.

We will now argue how this requirement translates in the identification of the space of gauge inequivalent holomorphicity preserving  infinitesimal deformations of a generalized complex cycle $(\Sigma,\calf)$ with an appropriate cohomology class on $(\Sigma,\calf)$.

Given a generalized complex cycle $(\Sigma,\calf)$, let us define on $\Sigma$ the holomorphic generalized tangent bundle $L_{(\Sigma,\calf)}=L|_{\Sigma}\cap(T_{(\Sigma,\calf)}\otimes \mathbb{C})$.
As explained in \cite{kapustindeform}, $L_{(\Sigma,\calf)}$ (as a complex vector bundle on $\Sigma$) can be equipped with a Lie algebroid structure as follows. The anchor map $\pi:L_{(\Sigma,\calf)}\rightarrow T_\Sigma\otimes \mathbb{C}$ is the obvious projection to $T_\Sigma\otimes \mathbb{C}$. The Lie brackets are defined by extending the sections of $L_{(\Sigma,\calf)}$ outside $\Sigma$ as sections of $L$, taking their Courant brackets,
and then restricting back to $\Sigma$ (and the result does not depend on the extension).

Thus, one can consider the exterior algebra bundle (see for example \cite{gualtieri})
\bea
 \Lambda^\bullet L^\star_{(\Sigma,\calf)}\equiv\bigoplus_{k}\Lambda^k L^\star_{(\Sigma,\calf)}
\eea
and a degree-1 derivation $d_{L_{(\Sigma,\calf)}}$ on its sections
\bea
d_{L_{(\Sigma,\calf)}}=\Gamma(\Lambda^k L^\star_{(\Sigma,\calf)})\rightarrow \Gamma(\Lambda^{k+1} L^\star_{(\Sigma,\calf)})\ .
\eea
Lie algebroid cohomology groups can then be naturally defined as
\bea
H^k(L_{(\Sigma,\calf)})=\frac{{\rm ker}[d_{L_{(\Sigma,\calf)}}:\Gamma(\Lambda^{k}L^\star_{(\Sigma,\calf)})\rightarrow \Gamma(\Lambda^{k+1}L_{(\Sigma,\calf)})]}{{\rm im}[d_{L_{(\Sigma,\calf)}}: \Gamma(\Lambda^{k-1}L^\star_{(\Sigma,\calf)})\rightarrow \Gamma(\Lambda^{k}L_{(\Sigma,\calf)})]}\ .
\eea

It is now important to note that the natural metric $\cali$ defines a canonical isomorphism $\tilde{}: T^\star_{(\Sigma,\calf)} \rightarrow \caln_{(\Sigma,\calf)}$
that associates (fiberwise) an element $\alpha\in T^\star_{(\Sigma,\calf)}$ to a generalized normal vector $[\tilde\alpha]$ such that, for any $\mathbb{X}\in T_{(\Sigma,\calf)}$,
\bea
\alpha(\mathbb{X})=2\cali(\tilde\alpha, \mathbb{X})\ ,
\label{isoNT}
\eea
where the factor of $2$ has been introduced to simplify some of the following expressions and manipulations.
This isomorphism preserves the generalized complex structure, in the sense that it also implies that $\caln^{0,1}_{(\Sigma,\calf)}\simeq L^\star_{(\Sigma,\calf)}$ where $\caln^{0,1}_{(\Sigma,\calf)}\equiv \bar L |_\Sigma/\bar L_{(\Sigma,\calf)}$, and can  be trivially extended to an isomorphism between $\Lambda^k \caln^{0,1}_{(\Sigma,\calf)}$ and $\Lambda^k L^\star_{(\Sigma,\calf)}$.

Let us now introduce the space $\Omega^{0,\bullet}_{(\Sigma,\calf)}$ of generalized currents with support on $\Sigma$ of the form\footnote{\label{tclass}In
a sense this provides a refinement and a generalization to currents of the decomposition in eq.~\eqref{dec}. Indeed, instead of $\psi$ one can use
$j_{(\Sigma,\calf)}$ as the base pure spinor acted upon with normal vectors leading to the decomposition $J_k$. If one further separates the normal vectors
in $+i$- and $-i$-eigenvectors of $\calj$ one finds the refined decomposition $J_{p,q}$. Now, $\Omega^{0,q}_{(\Sigma,\calf)}$ is the space of
sections of $J_{0,q}$.}
\bea
\Omega^{0,k}_{(\Sigma,\calf)}=\Gamma(\Lambda^k \caln^{0,1}_{(\Sigma,\calf)})\cdot j_{(\Sigma,\calf)}\ .
\eea
$\Omega^{0,\bullet}_{(\Sigma,\calf)}$  is clearly isomorphic to  $\Gamma(\Lambda^\bullet L^\star_{(\Sigma,\calf)})$ and in particular it is easy to see that for any $\mathbb{X}\in\Gamma(L_{(\Sigma,\calf)})$
\bea\label{inner}
\mathbb{X}\cdot\tilde\alpha\cdot j_{(\Sigma,\calf)}=(\widetilde{\iota_{\mathbb{X}}\alpha})\cdot j_{(\Sigma,\calf)}
\eea

As it will be clear in a moment, $\bar\partial_H$ acts as a degree-1 differential on $\Omega^{0,\bullet}_{(\Sigma,\calf)}$:
\bea
\bar\partial_H:\Omega^{0,k}_{(\Sigma,\calf)}\rightarrow \Omega^{0,k+1}_{(\Sigma,\calf)}\ .
\eea
Thanks to the isomorphism $\tilde{}$ , we can now define  an isomorphism between $\Gamma(\Lambda^k L^\star_{(\Sigma,\calf)})$ and $\Omega^{0,k}_{(\Sigma,\calf)}$ as follows:
\bea
\alpha\in \Gamma(\Lambda^k L^\star_{(\Sigma,\calf)})\quad\leftrightarrow\quad [\tilde\alpha]\cdot j_{(\Sigma,\calf)}\in\Omega^{0,k}_{(\Sigma,\calf)}\ .
\eea

Now, the key observation is that the isomorphism between $\Gamma(\Lambda^\bullet L^\star_{(\Sigma,\calf)})$ and $\Omega^{0,\bullet}_{(\Sigma,\calf)}$ preserves  the action of the derivative operators $d_{L_{(\Sigma,\calf)}}$ and $\bar\partial_H$. This means that
\bea\label{diffrel}
d_{L_{(\Sigma,\calf)}}\beta\in \Gamma(\Lambda^\bullet L^\star_{(\Sigma,\calf)})\quad\leftrightarrow\quad  \bar\partial_H([\tilde\beta]\cdot j_{(\Sigma,\calf)})=[\widetilde{d_{L_{(\Sigma,\calf)}}\beta}]\cdot j_{(\Sigma,\calf)} \in\Omega^{0,\bullet}_{(\Sigma,\calf)}\ .
\eea
This result can be proven by showing that, given a $\beta\in\Gamma(\Lambda^k L^\star_{(\Sigma,\calf)})$,
\bea
\mathbb{X}_1\cdot\ldots\cdot\mathbb{X}_{k+1}\cdot (\widetilde{d_{L_{(\Sigma,\calf)}}\beta})\cdot j_{(\Sigma,\calf)}=\mathbb{X}_1\cdot\ldots\cdot\mathbb{X}_{k+1}\cdot \bar\partial_H(\tilde\beta\cdot j_{(\Sigma,\calf)})\ ,
\eea
for any $\mathbb{X}_1,\ldots,\mathbb{X}_{k+1}\in\Gamma(L_{(\Sigma,\calf)})$.
The equality above can be checked by induction, using the definitions of $d_{L_{(\Sigma,\calf)}}\beta$ and $\bar\partial_H$, and the fact that from Lemma 4.24 of \cite{gualtieri} one has the following identity\footnote{
This identity is immediate when defining the Courant bracket as a derived bracket $[\mathbb{X},\mathbb{Y}]_H \cdot = [\call_{\mathbb{X}}, \mathbb{Y} \cdot ]$ \cite{derived,grananil}.}
\bea
\mathbb{X}_1\cdot\mathbb{X}_2\cdot d_H(\tilde\beta\cdot j_{(\Sigma,\calf)})&=&d_H(\mathbb{X}_2\cdot\mathbb{X}_1\cdot\tilde\beta\cdot j_{(\Sigma,\calf)})+\mathbb{X}_2\cdot d_H(\mathbb{X}_1\cdot\tilde\beta\cdot j_{(\Sigma,\calf)})\cr &&-\mathbb{X}_1\cdot d_H(\mathbb{X}_2\cdot\tilde\beta\cdot j_{(\Sigma,\calf)})+[\mathbb{X}_1,\mathbb{X}_2]_H\cdot\tilde\beta\cdot j_{(\Sigma,\calf)}\ ,
\eea
for any $[\tilde\beta]\in\Gamma(\Lambda^\bullet\caln_{(\Sigma,\calf)}^{0,1})$ and  any $\mathbb{X}_1,\mathbb{X}_2\in\Gamma(L_{(\Sigma,\calf)})$.

From (\ref{diffrel}) one can immediately conclude that the Lie algebroid cohomology groups $H^k(L_{(\Sigma,\calf)})$ are isomorphic
to the current cohomology groups
\bea
H^k_{\bar\partial_H}(\Sigma,\calf)=\frac{{\rm ker}(\bar\partial_H:\Omega^{0,k}_{(\Sigma,\calf)}\rightarrow\Omega^{0,k+1}_{(\Sigma,\calf)})}{{\rm im}(\bar\partial_H:\Omega^{0,k-1}_{(\Sigma,\calf)}\rightarrow\Omega^{0,k}_{(\Sigma,\calf)})}\ .
\eea

A fine point is that to properly define the Courant bracket we need to extend the sections of $L_{(\Sigma,\calf)}$
off the D-brane. The above cohomology groups do not depend on the extension (as long as they are still sections
of $L$ of course). Normally, we don't need to worry about this subtle issue unless we are working in points where the type
of the generalized complex structure changes. We will provide an example in section \ref{CY5}.

Consider now a generalized normal vector field $[\mathbb{X}_\lambda]$ which generates a real world-volume gauge transformation.
As we have already explained,  a representative of $[\mathbb{X}_\lambda]$ is given by $\mathbb{X}_\lambda=\xi$ with $\xi$ any 1-form such that $P_\Sigma[\xi]=d\lambda$. For our purposes we can extend the world-volume function $\lambda$ outside $\Sigma$ and simply pose $\mathbb{X}_\lambda=d\lambda$, and of course the following discussion will not depend on the way $\lambda$ is extended. Thus we have
\bea
\mathbb{X}_\lambda\cdot j_{(\Sigma,\calf)}=d\lambda\wedge j_{(\Sigma,\calf)}=d_H(\lambda j_{(\Sigma,\calf)})\ .
\eea
Since $j_{(\Sigma,\calf)}\in \Gamma_{\rm curr}(U_0)$, we can immediately conclude that
\bea\label{gaugedec}
\mathbb{X}^{0,1}_\lambda\cdot j_{(\Sigma,\calf)}=\bar\partial_H(\lambda j_{(\Sigma,\calf)})\ .
\eea

Since we have to identify a generalized normal vector field $[\mathbb{Y}^{0,1}]\in\Gamma(\caln_{(\Sigma,\calf)}^{0,1})$ with $[\mathbb{Y}^{0,1}]+[\mathbb{X}^{0,1}_\lambda]$, we see from (\ref{defcond}) and (\ref{gaugedec}) that the gauge inequivalent first order deformations of a generalized cycle $(\Sigma,\calf)$ are given by
\bea\label{firstcoho}
H^1_{\bar\partial_H}(\Sigma,\calf)\simeq H^1({L_{(\Sigma,\calf)}})\ .
\eea

\section{Infinitesimal deformations of generalized calibrated cycles}
\label{Dterm}

In section \ref{deformations} we have derived the condition (\ref{defcond}) that an infinitesimal deformation
generated by $\mathbb{X}$ must satisfy in order to deform a generalized complex cycle into another generalized complex cycle.
Furthermore we saw in section \ref{gauge} how considering infinitesimal deformations related by a {\em complexified} gauge transformation as equivalent leads to the  first cohomology group (\ref{firstcoho}) as the natural tangent space to the moduli space of inequivalent generalized complex cycles. In this section we will see how these equivalence classes of infinitesimal deformations corresponds to infinitesimal deformations of the generalized {\em calibrated} cycles of \cite{koerber,lucal}.

In \cite{koerber,lucal} it was shown how type II superstring backgrounds with fluxes can be characterized in terms of appropriate generalized calibrations, which `calibrate' the possible supersymmetric D-branes.
In particular, \cite{koerber} considered backgrounds with general pure NSNS fluxes on an internal space of arbitrary dimension,
leading to an $\caln=2$ residual supersymmetry in the external flat directions.
On the other hand \cite{lucal,branesuppot} focused on the class of so called {\em D-calibrated} backgrounds
with internal six-dimensional manifolds and arbitrary non-trivial NSNS and RR fluxes, preserving $\caln=1$ supersymmetry
in the flat four dimensions. The D-calibrated backgrounds constitute the most general class of $\caln=1$ backgrounds admitting `in principle' supersymmetric D-branes and can be seen as a  subclass of the vacua considered in \cite{gmpt}.

Since we want to consider backgrounds with possible minimal supersymmetry and general internal fluxes switched on, we will focus on the case studied in \cite{lucal,branesuppot} where $M$ has $d=6$. The results of \cite{koerber} when $d=6$  can be obtained as a subcase, by multiplying in an obvious way the appropriate quantities by an arbitrary constant phase coming from the underlying $\caln=2$ supersymmetry. The following results should also be straightforwardly extendible to the cases with $d\neq 6$ and arbitrary RR fluxes turned on.\footnote{ See \cite{witt3} for a proposal to extend the ideas of \cite{koerber,lucal} to different generalized geometries.}

Now, in \cite{lucal} it was shown how the generalized calibrations for D-branes on the $\caln=1$ backgrounds considered in \cite{gmpt} can be naturally written in terms of the two pure spinors $\Psi^\pm$ characterizing the internal geometry.\footnote{Note that the same pure spinors are indicated with $\Phi_\pm$ in \cite{gmpt} and, like in \cite{lucal}, we use the supergravity conventions of \cite{fermions} (see \cite{lucal} for the explicit relation with the conventions of \cite{gmpt}).} Note that on these $\caln=1$ backgrounds,  the generalized calibrations {\em depend} on the number of flat directions filled by the D-brane  (while in the $\caln=2$ case considered in \cite{koerber} the flat directions are not relevant like in the standard Calabi-Yau case). More precisely we have different generalized calibrations for space-time filling D-branes, D-brane domain walls and D-brane strings (i.e. D-branes filling four, three or two flat space-time directions).
The existence of the domain-wall calibration is essentially equivalent to the existence of a $d_H$-closed pure spinor
\bea
\psi\equiv e^{3A-\Phi}\hat\Psi_2\ ,
\eea
 where (see \cite{lucal,branesuppot} for more details)
\bea
\hat\Psi_2=\frac{-8i}{|a|^2}\Psi^+\ {\rm for\ IIA}\quad,\quad \hat\Psi_2=\frac{-8i}{|a|^2}\Psi^-\ {\rm for\ IIB}\ ,
\eea
$\Phi$ is the dilaton and $e^{2A}$ is the warp factor multiplying the four-dimensional flat metric.
Thus, as we have recalled in subsection \ref{sub2}, the internal manifold $M$ has a natural (integrable) generalized complex structure $\calj=\calj_2$
associated to $\hat{\Psi}_2$ or, more strictly, $M$ is a generalized Calabi-Yau \`a la Hitchin \cite{hitchin}. When, in what follows, we decompose forms and
exterior derivatives with respect to a generalized complex structure as in section \ref{sub2}, it will be this one.

On the other hand the other pure spinor $\hat\Psi_1$ defined as
\bea
\hat\Psi_1=\frac{-8i}{|a|^2}\Psi^-\ {\rm for\ IIA}\quad,\quad \hat\Psi_1=\frac{-8i}{|a|^2}\Psi^+\ {\rm for\ IIB}\ ,
\eea
is {\em not} $d_H$-closed due to the presence of the RR fluxes and thus defines a (generically non-integrable) generalized {\em almost} complex structure $\hat\calj = \calj_1$.

In this paper we will focus on the deformations of generalized calibrated {\em space-time filling} D-branes
(as we have said, in the $\caln=2$ case considered in \cite{koerber} this restriction is not necessary
and everything we will say is automatically valid for the other cases).
Now, in \cite{lucal,branesuppot} it was discussed how for space-time filling D-branes the supersymmetry condition, which is equivalent to the generalized calibration condition,
can be decoupled in a pair of conditions which have a clear four-dimensional interpretation.  The first condition can be seen as an F-flatness condition (which comes from an appropriate superpotential, defined in section \ref{4d}) and is equivalent to the condition that the D-brane must wrap a generalized complex cycle $(\Sigma,\calf)$ with respect to $\calj_2$, as defined in section \ref{deformations}. The second condition can be seen as a D-flatness condition, and can be written in terms of the non-integrable pure spinor in the following way:
\bea\label{dflatness}
P_\Sigma[e^{2A-\Phi}{\rm Im}\,\hat\Psi_1]\wedge e^\calf|_{\rm top}=0\ .
\eea
Note that
\begin{equation}
d_H \left(e^{2A-\Phi}{\rm Im}\,\hat\Psi_1\right) =0\ ,
\label{imclosed}
\end{equation}
this polyform being the generalized calibration for the D-brane strings (see \cite{branesuppot} for a discussion on this relation).

The outcome of this brief review of the results of \cite{lucal,branesuppot} is that the infinitesimal deformations preserving the generalized calibration condition must separately preserve the F- and D-flatness condition. We already know that the preservation of the F-flatness condition boils down to the condition (\ref{defcond}) for the generator $\mathbb{X}$.
On the other hand, one can see from a calculation like in \eqref{varj} that $\mathbb{X}$ preserves the D-flatness condition if and only if
\bea\label{dvar}
P_\Sigma[\call_\mathbb{X}(e^{2A-\Phi}{\rm Im}\,\hat\Psi_1)]\wedge e^\calf|_{\rm top}=\langle d_H(e^{2A-\Phi}\mathbb{X}\cdot{\rm Im}\,\hat\Psi_1),j_{(\Sigma,\calf)}\rangle =0\ ,
\eea
where the second term must be seen as a current density with support on $\Sigma$. One can show from \eqref{dflatness} that this condition is invariant
under the shift $\mathbb{X} \rightarrow \mathbb{X} + \mathbb{Y}$ with $\mathbb{Y}$ a section of $T_{(\Sigma,\calf)}$.
Since $j_{(\Sigma,\calf)}\in\Gamma_{\rm curr}(U_0)$, this condition can also be written in the form
\bea\label{realcond}
{\rm Re}\,\langle \partial_H(e^{2A-\Phi}\mathbb{X}^{0,1}\cdot{\rm Im}\,\hat\Psi_1),j_{(\Sigma,\calf)}\rangle =0\ .
\eea

The condition (\ref{realcond}) provides a gauge-fixing condition for the imaginary extension of the world-volume gauge transformations.
This was already argued in \cite{branesuppot} considering directly the condition (\ref{dflatness}) and showing that any imaginary gauge transformation violates it. The argument was based on the possibility of seeing (\ref{dflatness}) as the vanishing condition of the moment map associated to the real gauge transformations. We review the definition of the moment map in appendix \ref{mm}, in a form suited for the purposes of the present paper,
and apply it to also show how a possible violation of the condition (\ref{realcond}) under an infinitesimal deformation can always be reabsorbed by an appropriate imaginary gauge transformation. In short, in appendix \ref{mm} we show that in the equivalence class of deformations preserving the F-flatness condition there is one and only one
deformation preserving the D-flatness condition.

The same arguments of appendix \ref{mm} imply on the other hand that a condition of the form
\bea\label{imcond}
{\rm Im}\,\langle \partial_H(e^{2A-\Phi}\mathbb{X}^{0,1}\cdot{\rm Im}\,\hat\Psi_1),j_{(\Sigma,\calf)}\rangle =0\ ,
\eea
provides a gauge-fixing of the {\em real} gauge transformations. Indeed, it is enough to rewrite (\ref{imcond}) in the form
\bea
0&=&\langle d^{\calj}_H(e^{2A-\Phi}\mathbb{X}\cdot{\rm Im}\,\hat\Psi_1),j_{(\Sigma,\calf)}\rangle =\cr
&=&\langle d_H(e^{2A-\Phi}(\calj\mathbb{X})\cdot{\rm Im}\,\hat\Psi_1),j_{(\Sigma,\calf)}\rangle\ .
\eea
We can thus repeat the arguments in appendix \ref{mm} simply replacing $\mathbb{X}$ with $\calj\mathbb{X}$,
to see that (\ref{imcond}) selects a particular element in the equivalence class of real gauge transformations.

Putting together the two conditions (\ref{realcond}) and (\ref{imcond}) we obtain
\bea\label{complexcond}
\langle {\partial}_H(e^{2A-\Phi}\mathbb{X}^{0,1}\cdot{\rm Im}\,\hat\Psi_1),j_{(\Sigma,\calf)}\rangle =0\ ,
\eea
that provides a gauge-fixing for the whole complexified gauge algebra and thus identifies the $d_{L_{(\Sigma,\calf)}}$-closed  generator $\mathbb{X}^{0,1}$ as a particular element in the associated cohomology class.
What we find in the present context is what usually happens in the study of supersymmetric gauge theories,
where the flat directions in a (regular) point of the moduli space (in fact, the full moduli space, up to stability conditions)
can be described using holomorphic operators which are gauge invariant with respect to the complete {\em complexified} gauge group (see e.g. \cite{taylor}).

In fact, we can understand the above discussion by writing the condition (\ref{complexcond})
in a way which is the direct generalization of the usual conditions which are adopted in the standard case of fluxless D-branes on Calabi-Yau manifolds.
As discussed in appendix \ref{hodge}, it is possible to introduce a metric $G$ on the sections of $\Lambda^\bullet L^\star_{(\Sigma,\calf)} \simeq \Lambda^\bullet \caln_{(\Sigma,\calf)}^{0,1}$ and,
using it, to define a codifferential operator $d^\dagger_{L_{(\Sigma,\calf)}}$. The condition (\ref{complexcond}) can then be  rewritten in the following way:
\bea\label{hodgeD}
d^\dagger_{L_{(\Sigma,\calf)}}[\mathbb{X}^{0,1}]=0\ .
\eea

Recalling the F-flatness condition, we then find that  the supersymmetry preserving deformations are described by a section $[\mathbb{X}^{0,1}]$ of $\caln_{(\Sigma,\calf)}^{0,1}\simeq L^\star_{(\Sigma,\calf)}$ satisfying the conditions
\begin{equation}\label{adjoint}
d_{L_{(\Sigma,\calf)}} [\mathbb{X}]^{0,1}=0 \, , \qquad d_{L_{(\Sigma,\calf)}}^\dagger [\mathbb{X}]^{0,1}=0 \, .
\end{equation}
As an alternative to the proofs in appendix \ref{mm} one can thus also apply the standard formal argument to argue that the condition (\ref{hodgeD}) selects a unique element in the cohomology class represented by the $d_{L_{(\Sigma,\calf)}}$-closed element $[\mathbb{X}^{0,1}]$ (see for instance \cite{GSW}).
As usual, we can call such an element harmonic
\bea
\Delta_{L_{(\Sigma,\calf)}}[\mathbb{X}]^{0,1}=0\ ,
\eea
 with respect to  the generalized Laplacian
\bea\label{laplacian}
\Delta_{L_{(\Sigma,\calf)}}= d_{L_{(\Sigma,\calf)}} d_{L_{(\Sigma,\calf)}}^\dagger + d_{L_{(\Sigma,\calf)}}^\dagger d_{L_{(\Sigma,\calf)}}\ .
\eea

So it follows that the generalized calibration preserving deformations can be described by the harmonic representatives of the cohomology group $H^1(L_{(\Sigma,\calf)})$.
Furthermore, as discussed in appendix \ref{hodge}, one can see that the Lie algebroid differential complex associated to $L_{(\Sigma,\calf)}$ is elliptic and thus, since we always assume $\Sigma$ to be a compact cycle, its cohomology groups (and in particular $H^1(L_{(\Sigma,\calf)})$) are finite-dimensional.
Also, in appendix \ref{masses} we show how any deformation that preserves the minimal energy associated to the calibrated configuration must in fact
(after gauge-fixing the real gauge transformations) obey the conditions (\ref{adjoint}),
implying that $H^1(L_{(\Sigma,\calf)})$ indeed classifies the {\em massless} deformations around a calibrated cycle.

In summary, we have reached the conclusion that the infinitesimal gauge-inequivalent deformations preserving the generalized calibration condition for space-filling D-branes are given by elements of the following isomorphic cohomology groups
\bea
H^1_{\bar\partial_H}(\Sigma,\calf)\simeq H^1({L_{(\Sigma,\calf)}})\ ,
\eea
and in fact they are described by the harmonic representatives of these cohomology groups.

Note that, as a real  vector space, $H^1({L_{(\Sigma,\calf)}})$ has by construction a natural complex structure induced by $\calj$. This is expected from the effective four-dimensional description of the dynamics of the space-time filling D-branes,
that will be discussed more extensively in section \ref{4d}, since by supersymmetry
the massless fields should organize in complex chiral fields, whose number is then given by ${\rm dim}_\mathbb{C} \, H^1({L_{(\Sigma,\calf)}})$.

\section{Some comments on higher order deformations and  superpotentials}
\label{4d}

In the previous sections we have studied the first order deformations of generalized calibrated cycles
that preserve the calibration condition, finding that they are given by the elements of the first Lie algebroid cohomology group $H^1(L_{(\Sigma,\calf)})$.
The question of which of these first order deformations can actually be integrated to an unobstructed finite deformation is out of the scope of this paper.
The answer is bound to be non-trivial in general since it is known at least that in the extremal case of special Lagrangian manifolds there are no obstructions \cite{mclean}
while in the case of complex manifolds the obstructions lie in the first cohomology group ${H}^1_{\bar\partial}(\caln^{1,0}_\Sigma)$.\footnote{Actually, there is a case in which one can say something more. If we consider type IIA $SU(3)$ structure (symplectic) vacua, supersymmetric D6-branes must wrap Lagrangian cycles which satisfy an additional D-flatness condition which looks formally identical to the `speciality' condition for Lagrangian cycles in ordinary Calabi-Yaus \cite{lucal,branesuppot}. Then in this case, as observed in \cite{marchesanoD6}, the arguments of \cite{mclean} still apply and the D6-brane massless modes have no higher order obstructions.}
In this section we make some qualitative observations based on the four-dimensional point of view, but let us first note that  
for the deformations of the bulk on the other hand, it has been shown that they are unobstructed in the case of a compact $H$-twisted generalized Calabi-Yau \cite{li}, although
it is not known whether this result extends to the case where RR-fluxes spoil the integrability of one of the generalized complex structures.

In the four-dimensional description of a space-time filling D-brane,  $\dim_\mathbb{C}  H^1(L_{(\Sigma,\calf)})$ corresponds to the number of complex massless chiral fields $\phi^i$ coming from the deformations of the D-brane in the internal manifold. Considering a single D-brane that is always in the classical-geometrical regime, the $\phi^i$'s constitute the only chiral fields in the low-energy four-dimensional description resulting from integrating out the massive KK-deformations that do not preserve the calibration condition.
Furthermore, the $\phi^i$'s are uncharged under the low-energy $U(1)$ gauge symmetry.
The existence of higher order obstructions for the first order deformations given by $H^1(L_{(\Sigma,\calf)})$ is thus on physical grounds expected  to be equivalent to the existence of a non-trivial effective superpotential $\calw_{\rm eff}(\phi)$ for the chiral fields $\phi^i$.

In the six-dimensional case, this superpotential can in principle be obtained from the geometrical superpotentials found in \cite{branesuppot},
which have generalized complex cycles as extrema\footnote{Here we neglect possible world-sheet instanton effects that can indeed be present.}.
In order to understand why the six-dimensional case is special, let us rederive their form in a direct alternative way, using the characterization of generalized complex cycles given in equation (\ref{currcond}).
Let $\psi=e^{3A-\Phi} \hat{\Psi}_2$ denote the $d_H$-closed pure spinor giving $M$ the generalized Calabi-Yau structure.
Then, in the case of $M$ six-dimensional and only in this case, the condition (\ref{currcond}) can be written in the form
\bea
\int_M\langle \psi,\mathbb{X}\cdot j_{(\Sigma,\calf)}\rangle=0\quad,\quad {\rm for\ any\ }\mathbb{X}\in\Gamma\big((T_M\oplus T^\star_M)|_\Sigma\big)\ .
\eea
Note that in the above condition we can actually consider $\mathbb{X}$ as an element of $\Gamma(\caln_{(\Sigma,\calf)})$.
The latter can be identified with the tangent bundle of the configuration space of generalized cycles $\calc$ at the `point' $(\Sigma,\calf)$. Thus, if we introduce the one-form $\theta$ on $\calc$ defined as
\bea
\theta([\mathbb{X}])=\int_\Sigma P_\Sigma[\mathbb{X}\cdot \psi]\wedge e^\calf\ ,
\eea
for any $[\mathbb{X}]\in T_{\calc}|_{(\Sigma,\calf)}\simeq \Gamma(\caln_{(\Sigma,\calf)})$, we can identify the space $\calc_{\rm hol}\subset \calc$ of generalized complex cycles by the condition
\bea
\theta|_{\calc_{\rm hol}}=0\ .
\eea

The one-form $\theta$ is closed and the easiest way to see this is by writing it as $\theta=d\calw$ on some contractible domain in $\calc$.
To define $\calw$, let us first fix a certain (arbitrary) generalized cycle $(\Sigma_0,\calf_0)$.
Next, consider any cycle  $(\Sigma,\calf)$ in the same {\em generalized homology class}, i.e.\ this cycle should be
such that a generalized chain $(\calb,\tilde\calf)$ exists so that $\partial\calb=\Sigma-\Sigma_0$, $P_\Sigma[\tilde\calf]=\calf$ and $P_{\Sigma_0}[\tilde\calf]=\calf_0$.
Then the superpotential at the `point' $(\Sigma,\calf)$ is given by\footnote{To simplify the expressions we normalized all factors to one. The appropriate normalizations are discussed in \cite{branesuppot}.}
\bea\label{supot}
\calw(\Sigma,\calf)=\int_\calb P_\calb[\psi]\wedge e^{\tilde\calf}\ .
\eea
$\calw$ represents the geometrical D-brane superpotential that was first derived in \cite{branesuppot}.
Therein it was also shown how it can be obtained from a more physical argument relating it to the tension of BPS domain walls,
by using the domain wall calibrations of \cite{lucal}. Note that it obviously only depends on the integrable generalized complex structure, where on the other
hand, as we argue in appendix \ref{k}, the K\"ahler potential also depends on the non-integrable one. 

As we have explained  at the beginning of section \ref{gauge},
the space of generalized complex cycles $\calc_{\rm hol}$ is preserved
by the action of the algebra of complexified infinitesimal gauge transformations ${\bf g}^{\mathbb{C}}$,
which we obtained by complexifying the world-volume gauge transformations.
As discussed in \cite{branesuppot} in the case of $\caln=1$ flux vacua,
using the $SU(3)\times SU(3)$ structure of the background one can define an almost complex structure on the complete configuration space $\calc$
that under restriction to $\calc_{\rm hol}$ reduces to the almost complex structure implicitly introduced in section \ref{deformations}. The superpotential $\calw$ is holomorphic with respect to this almost complex structure and is automatically invariant under the complexified gauge algebra  ${\bf g}^{\mathbb{C}}$.

As was recalled in section \ref{Dterm}, supersymmetric generalized cycles are precisely the generalized calibrated cycles and
must not only extremize the superpotential (\ref{supot}),
but also obey the D-flatness condition. Postponing to appendix \ref{stability} a discussion on the stability problem, i.e.\ the problem of whether the D-flatness condition can in fact be satisfied at all, let us assume that it is indeed fulfilled (this seems plausible if we consider only configurations in a neighbourhood of a stable/supersymmetric one). Thus, as usual in standard supersymmetric gauge theories, the D-flatness condition provides a slice of the imaginary extension of the gauge group,
so that the moduli space $\calm$ of supersymmetric/calibrated generalized cycles can be written as
\bea
\calm\equiv (\calc_{\rm hol}\cap \{{\rm D-flat\ generalized\ cycles}\})/\calg=\calc_{\rm hol}/\calg^\mathbb{C}\ ,
\eea
where $\calg$ and $\calg^{\mathbb{C}}$ denote the space of finite gauge transformations generated by ${\bf g}$ and its complexification ${\bf g}^\mathbb{C}$ respectively.

One should thus be able to obtain the effective superpotential $\calw_{\rm eff}(\phi)$ by expanding (\ref{supot}) and
integrating out the massive fields.
For example, a second order expansion of the superpotential $\calw$ around a generalized complex cycle $(\Sigma,\calf)\in\calc_{\rm hol}$
can be obtained from the discussion of section \ref{deformations}, and is given by
\bea\label{quadratic}
\nabla_{[\mathbb{X}]} \nabla_{[\mathbb{Y}]}\calw|_{(\Sigma,\calf)}=\int_\Sigma P_\Sigma[\mathbb{X}\cdot d_H(\mathbb{Y}\cdot\psi)]\wedge e^\calf=\int_\Sigma P_\Sigma[\mathbb{X}^{0,1}\cdot \bar\partial_H(\mathbb{Y}^{0,1}\cdot\psi)]\wedge e^\calf\ ,
\eea
where the expression does not depend explicitly on the actual form of the covariant derivative $\nabla$ on $\calc$ since we are restricting to $\calc_{\rm hol}$ where $\nabla_{[\mathbb{Y}]}\calw|_{\calc_{\rm hol}}\equiv 0$. The quadratic term (\ref{quadratic}) should give the propagator one should use in order to integrate out the massive modes to produce $\calw_{\rm eff}(\phi)$, as for example discussed in \cite{wittenCS}. This procedure is non-trivial and involves the use of the complete $SU(3)\times SU(3)$ structure to fix the gauge symmetry and perform the actual expansion and reduction of $\calw$, but we will not try to address this question here.


\section{Examples and applications}
\label{examples}

Presently, we want to apply the results of the previous sections to study the massless spectrum of supersymmetric space-time filling D-branes
in several $SU(3)\times SU(3)$ structure type II backgrounds preserving $\caln=2$ or $\caln=1$ four-dimensional supersymmetry.
The $\caln=2$ case includes backgrounds where the internal is space is an ordinary Calabi-Yau or a generalized K\"ahler space
with non-trivial NSNS-fluxes. The $\caln=1$ case involves the inclusion of RR-fluxes.
Note that, differently from the Calabi-Yau or generalized K\"ahler case, the properties of the $\caln=1$ backgrounds are specific to a six-dimensional internal space and cannot be easily extrapolated to different dimensions.
Furthermore, in the $\caln=1$ case, there is no arbitrary phase in the non-integrable pure spinor,
so the Fayet-Iliopoulos term $\xi$, which depends on it, cannot always be set to zero independently of the kind of D-brane. Thus, in  $\caln=1$ vacua we should consider a restricted class of supersymmetric D-branes, having the same supersymmetry `phase', as specified by the non-integrable pure spinor. For example, in the type IIB $SU(3)$ structure case, this nothing but the statement that space-filling supersymmetric D3- and D7-branes exist in the so-called type B backgrounds, while D5- and D9-branes exist in the type C backgrounds \cite{granareview,branesuppot}.
However, the key result of the previous sections is that, once we start from a supersymmetric D-brane configuration,  the massless fluctuations around it are given by the Lie algebroid cohomology $H^1(L_{(\Sigma,\calf)})$ which depends only on the integrable generalized complex structure, always present in both the $\caln=1$ and $\caln=2$ case.
The fact that the other pure spinor is integrable only in the $\caln=2$ case does not affect the above result on the spectrum of massless D-brane fluctuations.  Thus we will generically  consider our internal space as being simply a generalized Calabi-Yau space (\`a la Hitchin) of arbitrary dimension, restricting to the six-dimensional case when we need to be more specific.

We start with the ordinary Calabi-Yau case,
which is quite well understood and will be used here as a warm-up exercise,
and later we introduce world-volume and background fluxes.
When possible, in the cases where the fluxes  `preserve' an underlying complex or symplectic geometrical structure,
we discuss in detail the flux-induced moduli-lifting mechanism.  We also consider point-like cycles in a genuine generalized complex background, which even if seeming so simple, exhibit some interesting non-trivial features due to the generalized nature of the underlying geometry.

\subsection{D-branes on ordinary Calabi-Yaus: introduction}
\label{CY1}
The standard examples of generalized complex submanifolds are B-branes and A-branes in Calabi-Yau manifolds.  The former correspond to
complex submanifolds on which a holomorphic connection is defined, while the latter corresponds to the coisotropic branes of \cite{kapustincoisotropic,kapustinstability}. The  cohomology groups $H^k(L_{(\Sigma,\calf)})$ in these cases have already
been studied in detail in \cite{kapustindeform}, where they arose as the BRST cohomology giving the massless spectrum of the open topological string. Our result that identifies $H^1(L_{(\Sigma,\calf)})$ with the fluctuations of the calibrated generalized cycles provides the geometrical counterpart of the results of \cite{kapustindeform}.

In the following two subsections we will revisit,
for completeness and also as a useful warm-up exercise,
the essential points of the calculation of $H^k(L_{(\Sigma,\calf)})$ given in \cite{kapustindeform}.
This will allow us to point out some important observations, which will be useful in the subsequent subsections
where we consider D-branes on backgrounds with fluxes.

For B- and A-branes on Calabi-Yaus, it is enough to consider the Calabi-Yau's complex or symplectic structure respectively. We  thus recall here the form of  the corresponding globally defined pure spinors, $\psi_c$ and $\psi_s$, and  generalized complex structures $\calj_c$ and $\calj_s$.

From an almost complex structure $J$
one can construct a generalized  almost  complex structure
\begin{equation}\label{gcsc}
\calj_c = \left( \begin{array}{cc} -J & 0 \\ 0 & J^T \end{array}\right) \, ,
\end{equation}
with pure spinor of the form $\psi_c = \Omega$, where $\Omega$ is the holomorphic $(3,0)$-form associated to $J$. In the fluxless Calabi-Yau case $H=0$, and so $\calj_c$ is  integrable if and only if the complex structure $J$ is integrable.

From a non-degenerate antisymmetric form $\omega$ one
can construct a second type of generalized almost complex structure
\begin{equation}\label{gcss}
\calj_s = \left( \begin{array}{cc} 0 & \omega^{-1} \\ -\omega & 0 \end{array}\right) \, ,
\end{equation}
with pure spinor of the form $\psi_s = e^{i \omega}$. This generalized complex structure is  integrable if and only if $d \omega = 0$, and thus $\omega$ defines a symplectic structure.


\subsection{Warm-up:  fluxless supersymmetric A- and B-branes on Calabi-Yaus}
\label{fluxlessCY}

Let us start with the simplest and well-known case of zero world-volume flux $\calf$
(which is possible since $H=0$). Thus the wrapped cycle $\Sigma$ must be special Lagrangian for A-branes and holomorphic for B-branes.
The sector of the spectrum originating from the geometrical fluctuations of the cycles is given by $H^1(\Sigma,\mathbb{R})$ for A-branes \cite{mclean} and $H^0_{\bar\partial}(\caln^{1,0}_\Sigma)$ for B-branes, where $\caln^{1,0}_\Sigma$ is the (ordinary) holomorphic normal bundle to the brane.
In addition to these geometrical fluctuations, we must also consider the world-volume gauge field fluctuations.
Since the A-branes must preserve the condition $\calf=0$, the spectrum of gauge field fluctuations is given by another $H^1(\Sigma,\mathbb{R})$,
that combines with the one of the geometrical fluctuations to give $H^1(\Sigma,\mathbb{C})$.
On the other hand, the requirement for B-branes is that the line bundle connection must be holomorphic, so that the corresponding spectrum is given by $H^{0,1}_{\bar\partial}(\Sigma)$.
Thus, around $\calf=0$, the total B-brane deformations are given by  $H^0_{\bar\partial}(\caln^{1,0}_\Sigma)\oplus H^{0,1}_{\bar\partial}(\Sigma)$.

These spectra are directly reproduced by $H^1(L_{(\Sigma,\calf=0)})$, which automatically encodes information about both the geometrical and gauge sectors.
To show this, let us review explicitly the computation of $H^k(L_{(\Sigma,\calf=0)})$, first presented in \cite{kapustindeform}.

\smallskip

Consider first Lagrangian A-branes, i.e.\ branes that wrap middle-dimensional cycles $\Sigma$, such that $P_\Sigma[\omega]=0$. The background maximal isotropic sub-bundle defining the generalized complex structure (\ref{gcss}) is given by
\bea
L^s=\{X-i \, \iota_X\omega: X\in T_M\otimes \mathbb{C} \}\ ,
\eea
while the generalized tangent bundle is given by
\bea\label{gtbs}
T_{(\Sigma,\calf=0)}=T_{\Sigma}\oplus \caln^\star_\Sigma\ ,
\eea
where $\caln^\star_\Sigma\equiv {\rm Ann} \, T_\Sigma$ is the cycle's conormal bundle, dual to $\caln_\Sigma$. Thus the D-brane Lie algebroid is
\bea
L_{(\Sigma,\calf=0)}=L^s|_\Sigma\cap ( T_{(\Sigma,\calf=0)}\otimes \mathbb{C})=\{X-i \, \iota_X\omega: X\in T_\Sigma\otimes \mathbb{C} \}\ ,
\eea
and
\bea\label{agnor}
L^\star_{(\Sigma,\calf=0)}\simeq \caln_{(\Sigma,\calf=0)}^{0,1} =\{[V]+iP_\Sigma[\iota_V\omega]:[V]\in\caln_\Sigma\}\ .
\eea
Using the property \eqref{propcourant} of Courant brackets (with $B=-i\omega$) we can easily see that $L_{(\Sigma,\calf=0)}\simeq T_\Sigma\otimes \mathbb{C}$ not only as a bundle but also as a Lie algebroid on $\Sigma$.
Thus  $L^\star_{(\Sigma,\calf=0)}\simeq T^\star_\Sigma\otimes \mathbb{C}$, as can also be seen directly from (\ref{agnor}), and the Lie algebroid differential is given by the ordinary differential acting on $\Gamma(\Lambda^k T^\star_\Sigma)$, i.e.
\bea
d_{L_{(\Sigma,\calf=0)}} \simeq d\ .
\eea
For the sake of comparison to \cite{mclean} we explicitly construct the isomorphism $L^\star_{(\Sigma,\calf=0)}\simeq T^\star_\Sigma\otimes \mathbb{C}$.
We take $\tilde{\alpha}^{0,1}=([V]+iP_\Sigma[\iota_V\omega]) \in \caln_{(\Sigma,\calf=0)}^{0,1}$ and associate to this normal vector
in the manner of \eqref{isoNT} an element $\alpha \in L^\star_{(\Sigma,\calf)}$ that acts on $\mathbb{X}=X-i \, \iota_X\omega$
as
\begin{equation}
\alpha(\mathbb{X}) = 2 i \, \iota_X \iota_V \omega \, .
\end{equation}
The isomorphism is now given by associating to this a $\beta \in T^{\star}_\Sigma\otimes \mathbb{C}$
\begin{equation}
\beta = 2i \, \iota_V \omega \, .
\end{equation}

A massless deformation is thus described by a $V \in \Gamma(N_\Sigma)$ such that $\iota_V \omega$ is closed on $\Sigma$, but not exact.
Furthermore we find that exactly when $V$ is real, $\tilde{\alpha}^{1,0}+\tilde{\alpha}^{0,1}=2[V]$ describes a purely geometric deformation, i.e.\
one without deformation of the gauge field. In this case we find precisely the result of \cite{mclean}.

Concluding  $H^k(L_{(\Sigma,\calf=0)}) \simeq H^k(\Sigma,\mathbb{C})$, and in particular the massless deformations of special Lagrangian D-branes are given by  $H^1(\Sigma,\mathbb{C})$, reproducing the well-known result we recalled above.

\smallskip

Let us now consider B-branes wrapping holomorphic cycles with $\calf=0$.  In this case the relevant maximal isotropic sub-bundle is given by
\bea
L^c=T^{0,1}_M\oplus T^{\star 1,0}_M\ ,
\eea
while the generalized tangent bundle is still given by (\ref{gtbs}). Thus the Lie algebroid becomes
\bea
L_{(\Sigma,\calf=0)}=T_\Sigma^{0,1}\oplus \caln_\Sigma^{\star 1,0}\ ,
\eea
with Lie algebroid bracket given by
\bea\label{brkrules}
[X_1,X_2]_{L_{(\Sigma,\calf=0)}}=[X_1,X_2]\, ,\quad [X_1,\xi_1]_{L_{(\Sigma,\calf=0)}}=\iota_{X_1}d\xi_1\, ,\quad [\xi_1,\xi_2]_{L_{(\Sigma,\calf=0)}}=0\ ,
\eea
where $X_1,X_2\in\Gamma(T_\Sigma^{0,1})$ and $\xi_1,\xi_2\in\Gamma( \caln_\Sigma^{\star 1,0})$, while the dual $(0,1)$ generalized normal bundle is given by
\bea
L_{(\Sigma,\calf=0)}^\star\simeq \caln_{(\Sigma,\calf=0)}^{0,1}= T_\Sigma^{\star 0,1}\oplus \caln_\Sigma^{1,0}\ .
\eea
Thus the Lie algebroid differential complex is given by the sections of
\bea\label{complex1}
\Lambda^k L_{(\Sigma,\calf=0)}^\star\simeq \Lambda^k\caln_{(\Sigma,\calf=0)}^{0,1}=\bigoplus_{k=p+q}\Lambda^pT_\Sigma^{\star 0,1}\otimes \Lambda^q\caln_\Sigma^{1,0}\ ,
\eea
with differential
\bea
d_{L_{(\Sigma,\calf=0)}} \simeq \bar\partial\ ,
\eea
as can easily be computed from (\ref{brkrules}). The conclusion is that the algebroid cohomology is given by
\bea\label{cohoB1}
H^k(L_{(\Sigma,\calf=0)})=\bigoplus_{k=p+q}H^{0,p}_{\bar\partial}(\Lambda^q\caln^{1,0}_\Sigma)\ ,
\eea
and the massless deformations of supersymmetric B cycles with $\calf=0$ are given by
\bea
H^1(L_{(\Sigma,\calf=0)})=H^0_{\bar\partial}(\caln^{1,0}_\Sigma)\oplus H^{0,1}_{\bar\partial}(\Sigma)\ ,
\eea
reproducing the well-known result announced above.

\subsection{A less trivial case:  D-branes on Calabi-Yaus with world-volume flux}
\label{CY2}

In this subsection we consider the effect of introducing world-volume fluxes
on supersymmetric D-branes on Calabi-Yau manifolds.
The calculation of the relevant algebroid cohomology groups has already been performed in some detail in \cite{kapustindeform},
and here we recall only the main points. We will then discuss the interpretation of the resulting massless fluctuations
of the supersymmetric D-branes, trying to clarify the role of fluxes and their possible moduli lifting effect.
Note that the effect of world-volume fluxes on D-branes in Calabi-Yau manifolds has been discussed
from a geometrical point of view several times in the literature (see for example \cite{mmms,leung1,leung2,sharpe1,sharpe2}).
Here we would like to revisit this issue from the point of view of generalized complex geometry.

\subsubsection{B-branes with flux}
\label{Bflux}

Let us start with supersymmetric B-branes with world-volume fluxes, which are more intensively studied than the flux A-branes.
Thus, the D-brane wraps a generalized cycle $(\Sigma,\calf)$ that is generalized complex with respect to $\calj_c$.
According to \cite{gualtieri} this means that $\Sigma$ is a complex submanifold with respect to $J$,
and $\calf$ is of type $(1,1)$ with respect to $P_\Sigma[J]$. So we end up with a holomorphic cycle on which a holomorphic line bundle is defined.

Following the same steps as in the previous examples, one can easily  find that $L_{(\Sigma,\calf)}$ is spanned by vectors of the form
\begin{equation}
\begin{split}
\label{complexvectors}
\mathbb{X}_1 & = \bar X + \iota_{\bar X} \calf  \, , \qquad \bar X \in T_\Sigma^{0,1} \, , \\
\mathbb{X}_2 & =  \eta  \, , \qquad \qquad \quad\,\,\, \eta \in \caln_\Sigma^{\star 1,0} \, .
\end{split}
\end{equation}
At a first sight, the identification of the vectors (\ref{complexvectors}) seems to provide a possible splitting
of $L_{(\Sigma,\calf)}$ into
\begin{equation}\label{splitting}
T_\Sigma^{0,1} \oplus \caln_\Sigma^{\star 1,0} \, .
\end{equation}
This would give an isomorphism $L_{(\Sigma,\calf)}\simeq L_{(\Sigma,\calf=0)}$
that would furthermore be compatible with the Lie algebroid brackets,
leading to the conclusion that the algebroid cohomology would still be given by  (\ref{cohoB1}).

However, it is in general not possible to canonically split  $L_{(\Sigma,\calf)}$ as in (\ref{splitting}), since the presence of a non-zero $\calf$ generically implies that the transition functions on overlapping patches will mix vectors of type $\mathbb{X}_1$ and $\mathbb{X}_2$.
The obstruction to such a splitting originates in an obstruction to the splitting of $T_M^{1,0}|_\Sigma$
into $T_\Sigma^{1,0}\oplus \caln_\Sigma^{1,0}$ as holomorphic bundles.

To see this, we can take the dual bundle $T_M^{\star 1,0}|_\Sigma$,  that analogously does not generically allow the holomorphic splitting
\bea\label{holsplit}
T_M^{\star 1,0}|_\Sigma\simeq_{\rm hol}T_\Sigma^{\star 1,0}\oplus \caln_\Sigma^{\star 1,0} \, ,
\eea
but instead fits into the short exact sequence
\bea
0\ \rightarrow\ \caln^{\star 1,0}_\Sigma\ \rightarrow\ T_M^{\star 1,0}|_\Sigma\ \rightarrow\  T_\Sigma^{\star 1,0}\ \rightarrow\ 0\ .
\eea
Thus we see that  the terms of the form $\iota_{\bar X} \calf\in T_\Sigma^{\star 1,0}$ in vectors of type  $\mathbb{X}_1$,
will generically mix with elements of  $\caln^{\star 1,0}_\Sigma$ that enter the vectors of type $\mathbb{X}_2$.
On the other hand, in the case that the holomorphic splitting (\ref{holsplit}) is allowed,
we can indeed  conclude that the algebroid cohomology groups $H^k(L_{(\Sigma,\calf=0)})$ are given by (\ref{cohoB1}),
and then there is no flux-induced moduli lifting mechanism and the spectrum of massless fluctuations is identical to the fluxless case.

Let us now describe the main points of  what happens when the holomorphic splitting is not allowed (for more details see \cite{kapustindeform}).  It is first convenient to find a consistent way to define the splitting of $L_{(\Sigma,\calf)}$ into (\ref{splitting}), so that $L_{(\Sigma,\calf)}$ and $L_{(\Sigma,\calf=0)}$ are given by the same vector bundle, but with different Lie algebroid structures.

This can be done by choosing a particular set of holomorphic charts adapted to $\Sigma$.
They define a set of holomorphic sections $\gamma$ of $T_\Sigma^{1,0}\otimes \caln_\Sigma^{\star 1,0}$
on the overlap of the different pairs of charts,
that by definition give the mixing terms in $\caln_\Sigma^{\star 1,0}$, produced by elements of  $T_\Sigma^{\star 1,0}$
when passing from one chart to another. This means that, if $U$ and $U^\prime$ are two intersecting charts adapted to $\Sigma$,
then a section $\eta$ of $T_\Sigma^{\star 1,0}|_U$ is on $U\cap U^\prime$ related by the change of charts to the element $\eta+\gamma\lrcorner \eta$.
$\gamma$ can easily be seen to define a \u{C}ech 1-cocycle with values in the holomorphic sheaf $T_\Sigma^{1,0}\otimes \caln_\Sigma^{\star 1,0}$
and $\bar{\partial} \gamma=0$.
The 1-cocycle $\gamma$ can be written as the \u{C}ech coboundary
of a smooth 0-cochain $p$ of  $T_\Sigma^{1,0}\otimes \caln_\Sigma^{\star 1,0}$:
\bea\label{exact}
\gamma=\delta p\ ,
\eea
i.e., $p$ is defined by smooth sections on the different charts and, if $U$ and $U^\prime$ are two intersecting charts, then $\gamma=p^\prime-p$ on $U\cap U^\prime$.

Then we have $L_{(\Sigma,\calf)}\simeq T_\Sigma^{0,1} \oplus \caln_\Sigma^{\star 1,0}$ defined by the inclusion  $\caln_\Sigma^{\star 1,0}\subset L_{(\Sigma,\calf)}$ and by identifying $\bar X\in T_\Sigma^{0,1}$ on each chart with the element
\bea
\bar X+\iota_{\bar X}\calf-p\lrcorner (\iota_{\bar X}\calf)\in  L_{(\Sigma,\calf)}\ .
\eea
Using the transformation laws $p \rightarrow p + \gamma$ and $\iota_{\bar X}\calf \rightarrow \gamma \lrcorner (\iota_{\bar X}\calf)$ under change of charts, one can check that the above splitting is well defined.

Now, $L_{(\Sigma,\calf)}$ and $ T_\Sigma^{0,1} \oplus \caln_\Sigma^{\star 1,0}$, even if isomorphic as vector spaces,
are not isomorphic as Lie algebroids.
This implies that the algebroid differential $d_{L_{(\Sigma,\calf)}}$ still acts on the smooth sections of (\ref{complex1})
but is different from $\bar\partial$.
To describe it, let us first note that since $\bar{\partial} \gamma=0$ we have $\delta \bar{\partial} p=0$
so that $\bar\partial p$ defines a $\bar\partial$-closed {\em global} section $\beta_\Sigma$ of $T_\Sigma^{1,0}\otimes \caln_\Sigma^{\star 1,0}\otimes T_\Sigma^{\star 0,1}$, which defines a cohomology class $[\beta_\Sigma]\simeq [\gamma]$ in $H^{0,1}_{\bar\partial}(T_\Sigma^{1,0}\otimes \caln_\Sigma^{\star 1,0})$. Then, it is possible to show that
\bea\label{difflux}
d_{L_{(\Sigma,\calf)}} \simeq \bar\partial+\delta(\Sigma,\calf)\ ,
\eea
with
\bea\label{beta}
\delta(\Sigma,\calf)\equiv -(\beta_\Sigma\circ \calf)\llcorner\,: \Gamma(\Lambda^pT^{\star 0,1}_M\otimes \Lambda^q\caln_\Sigma^{1,0})\rightarrow \Gamma(\Lambda^{p+2}T^{\star 0,1}_M\otimes \Lambda^{q-1}\caln_\Sigma^{1,0})\ ,
\eea
where
\bea
\circ: \, T_\Sigma^{1,0}\otimes \caln_\Sigma^{\star 1,0}\otimes T_\Sigma^{\star 0,1}\, \times \,  T_\Sigma^{\star 1,0}\otimes T_\Sigma^{\star 0,1}\, \rightarrow \, \caln_\Sigma^{\star 1,0} \otimes \Lambda^2T_\Sigma^{\star0,1}
\eea
is given by (the unique) combination of external product and contraction, and $\llcorner$  contracts the index in $\caln_\Sigma^{\star 1,0}$ of $\beta_\Sigma\circ \calf$ with the indices in $\Lambda^q\caln_\Sigma^{1,0}$.

Note that $\delta(\Sigma,\calf)$ in (\ref{difflux}) is itself a differential (of degree $(2,-1)$) and, since $\beta_\Sigma\circ \calf$ is $\bar\partial$-closed, $d_{L_{(\Sigma,\calf)}} $can be seen as the sum of two commuting differentials. The study of the Lie algebroid cohomology could be addressed by  application of the standard theory of spectral sequences. However, the calculation of $H^1(L_{(\Sigma,\calf)})$ giving the D-brane massless fluctuations can be performed directly. Indeed, a pair $(a,[X])\in \Gamma(L^\star_{(\Sigma,\calf)})\simeq\Gamma(T^{\star 0,1}_\Sigma \oplus \caln^{1,0}_\Sigma)$ defines an element in  $H^1(L_{(\Sigma,\calf)})$ if and only if
\bea
&& \bar \partial[X]=0\quad, \label{1} \\
&&\delta(\Sigma,\calf)\cdot [X]+ \bar\partial a=0\quad, \label{2} \\
&& a\simeq a+\bar\partial\lambda\quad {\rm for\ any\ } \lambda\in\Gamma(\Sigma,\mathbb{C})\ .\label{3}
\eea
The first condition (\ref{1}) tells us that $[X]\in H^0_{\bar\partial}(\caln_\Sigma^{1,0})$. Note that it also implies that $\delta(\Sigma,\calf)\cdot [X]$ is $\bar\partial$-closed, and the second condition (\ref{2}) imposes that it must be actually exact and equal to $-\bar\partial a$. Thus, $a$ is determined by (\ref{2}) up to a $\bar\partial$-closed term, and (\ref{3}) means that we must only consider the corresponding element in $H^{0,1}_{\bar\partial}(\Sigma)$.

Thus we reach the conclusion that the massless fluctuations are given by
\bea\label{fluxfluctuations}
H^1(L_{(\Sigma,\calf)})= H^{0,1}_{\bar\partial}(\Sigma)\oplus  H^0_{d_{L_{(\Sigma,\calf)}}}(\caln^{1,0}_\Sigma)\ ,
\eea
where
\bea\label{gmassless}
H^0_{d_{L_{(\Sigma,\calf)}}}(\caln^{1,0}_\Sigma)={\rm ker}[ \delta(\Sigma,\calf): H^0_{\bar\partial}(\caln^{1,0}_\Sigma)\rightarrow H^{0,2}_{\bar\partial}(\Sigma)]\ .
\eea

From  (\ref{fluxfluctuations}) we can immediately conclude that the massless modes originating from the world-volume gauge field (given by  $H^{0,1}_{\bar\partial}(\Sigma)$)  are not lifted by the flux $\calf$.
This was expected, since we can consider fluctuations of the holomorphic gauge field while keeping the cycle $\Sigma$ {\em fixed}, thus always obtaining  $H^{0,1}_{\bar\partial}(\Sigma)$ as part of the spectrum.
On the other hand, if the holomorphic splitting (\ref{holsplit}) is not allowed,
 the action of $\delta(\Sigma,\calf)$ on $H^0_{\bar\partial}(\caln^{1,0}_\Sigma)$ is in general non-trivial and the geometrical massless modes of the fluxless case can be lifted.

We can give a direct geometrical explanation of (\ref{2})  and of the resulting lifting of the geometrical massless modes encoded in (\ref{gmassless}).
Suppose first that $T^{1,0}_M|_{\Sigma}$ splits holomorphically into $T^{1,0}_{\Sigma}\oplus \caln^{1,0}_{\Sigma}$. Any holomorphic section  $[X]$ of $\caln_{\Sigma}^{1,0}$ can be uplifted to a globally defined holomorphic section of $T_M^{1,0}|_\Sigma$. Thus, $[X]$ generates a deformation of the holomorphic cycle $\Sigma$ while keeping its complex structure {\em fixed}. This means that $\calf_{0,2}$ remains zero under the deformation of the cycle and $(\Sigma,\calf)$  deforms to a generalized complex cycle.

Let us now consider the general case in which $T_M^{1,0}|_\Sigma$ {\em does not} split holomorphically into $T_\Sigma^{1,0}\oplus \caln^{1,0}_\Sigma$.  In this case the holomorphic section $[X]$ of $\caln_{\Sigma}^{1,0}$  is uplifted to a {\em smooth} section $X$ of $T_M^{1,0}|_\Sigma$, which, generically, is not holomorphic.
However, $\bar\partial X$ is a globally defined holomorphic $(0,1)$-form with values in $T_\Sigma^{1,0}$, that can be taken to be given by $\beta_\Sigma\llcorner X$, and defines a cohomology class $[\beta_\Sigma\llcorner X]$  in $H^{0,1}_{\bar\partial}(T^{1,0}_\Sigma)$.\footnote{$[\beta_\Sigma\llcorner X]$ is nothing but the element associated to $[X]$ by the extension map $H_{\bar{\partial}}^0(\caln_{\Sigma}^{1,0})\rightarrow H^{0,1}_{\bar\partial}(T^{1,0}_\Sigma)$ in the long exact sequence associated to the short exact sequence
\bea
0\rightarrow T^{1,0}_\Sigma \rightarrow T_M^{1,0}|_\Sigma \rightarrow \caln_{\Sigma}^{1,0}\rightarrow 0\ .
\eea
}
The element $[\beta_\Sigma\llcorner X]$ gives the infinitesimal deformation of the complex structure of the holomorphic cycle $\Sigma$ induced by $[X]$.
This deformation can induce a violation of the condition $\calf_{0,2}=0$ and thus the only $[X]$'s which do not get a mass are those for which
\bea
(\delta_{[X]}\calf)_{0,2}=-(\bar\partial X)\circ \calf=\delta(\Sigma,\calf) \cdot X\ ,
\eea
can be compensated by a corresponding gauge-field deformation $a$ such that $\bar\partial a=-(\delta_{[X]}\calf)_{0,2}$, thus reproducing (\ref{2}).

Considering more specifically the case of Calabi-Yau three-folds, we can see that such a flux induced moduli-lifting can happen only for divisors.
Indeed,  in the case of zero- and six-cycles, we have no tangent or normal bundle respectively, and so obviously $\delta(\Sigma,\calf)\equiv 0$.
In the two-cycle case, we also have $\delta(\Sigma,\calf)\equiv 0$, since $\beta_\Sigma\circ \calf\in \Gamma( \caln_\Sigma^{\star 1,0}\otimes \Lambda^2T_\Sigma^{\star 0,1})$
and thus vanishes identically. The same conclusion namely that the massless modes --- and more generally the full moduli space --- of two-cycles do not depend on the world-volume fluxes, could be reached directly from the superpotential (\ref{supot}) for generalized two-cycles $(\Sigma_2,\calf)$, which is given by
\bea
{\calw}(\Sigma_2,\calf)=\int_{{\cal B}_3}P_{{\cal B}_3}[\Omega]\ ,
\eea
and thus clearly does not depend on $\calf$.

Thus the only case in which such a non-trivial effect can take place is given by a generalized four-cycle $(\Sigma_4,\calf)$, and again it can be obtained from the corresponding superpotential
\bea\label{hol4}
{\calw}(\Sigma_4,\calf)=\int_{{\cal B}_5}P_{{\cal B}_5}[\Omega]\wedge \tilde\calf\ ,
\eea
as has already been described  in \cite{branesuppot}.
Furthermore,  from (\ref{hol4}) the holomorphic mass matrix for the lifted modes can also be computed, and is given by
\bea
m_{ij}^{\rm (hol)}&=&\partial_i\partial_j\calw=-\int_{\Sigma_4} P_{\Sigma_4}[\iota_{X_i}\Omega]\wedge [(\bar\partial X_j)\circ \calf]=\cr
&=&\int_{\Sigma_4} P_{\Sigma_4}[\iota_{X_i}\Omega]\wedge [\delta(\Sigma,\calf)\cdot X_j] \ ,
\eea
where the $[X_i]$ form a base of $H^0_{\bar{\partial}}(\caln_\Sigma^{1,0})$. Thus $m_{ij}=0$ whenever $[\delta(\Sigma,\calf)\cdot X_i]=0$ or $[\delta(\Sigma,\calf)\cdot X_j]=0$ as elements of $H^{0,2}_{\bar{\partial}}(\Sigma)$, which is in accordance with the above general discussion.

\subsubsection{Coisotropic A-branes}
\label{coisotropic}

In \cite{kapustincoisotropic,gualtieri} it is explained that a D-brane $(\Sigma,\calf)$
is generalized complex with respect to $\calj_s$, i.e.\ it is an A-brane, if and only if
\begin{enumerate}
\item $\Sigma$ is coisotropic i.e.\ $\omega^{-1} \xi \in T_\Sigma \, , \forall \xi \in \caln^\star_\Sigma$.
  It follows that
the symplectic orthogonal bundle $T_\Sigma^\perp$ lies within $T_\Sigma$. It also follows that $T_\Sigma^\perp$ is integrable.
\item $\iota_X \calf =0 \, , \forall X \in T^\perp_\Sigma$ i.e.\ $\calf$ descends to  $\calt_\Sigma\equiv T_\Sigma/T^\perp_\Sigma$.
\item $(\omega|_{\calt_\Sigma})^{-1} \calf|_{\calt_\Sigma}$ is an almost complex structure on $\calt_\Sigma$. In fact, in \cite{kapustincoisotropic} it was shown
that it is integrable. In the rest of this subsection if we use the holomorphic decomposition of ordinary vectors or forms, it will be with
respect to this complex structure.
\end{enumerate}
When $T_\Sigma = T_\Sigma^\perp$ we find that the submanifold $\Sigma$ is also isotropic. The second condition then says that $\calf=0$ and the third is vacuous.
This is the standard fluxless Lagrangian case.

It is immediate to show that $iP_\Sigma[\omega]+\calf$ is a  non-degenerate $(2,0)$ form  on $\calt_\Sigma$.
It follows that the complex dimension of $\calt_\Sigma$ must be even and thus we can deduce that $\text{dim} \,\calt_\Sigma= 4k$.
We have then $\text{dim} \, \Sigma=d/2 + 2k$ and $\text{dim} \, T_\Sigma^\perp = \text{dim} \, \caln_\Sigma = d/2-2k$. The cases
relevant for a 6-dimensional internal manifold are $k=0$ (special Lagrangian) and $k=1$ (coisotropic D8-branes).
From \eqref{currcond} we find immediately an equivalent but more simple formulation
of the condition for a coisotropic D-brane
\begin{equation}
\left(i P_\Sigma[\omega]+\calf\right)^{k+1} = 0 \, .
\label{coisotropiccond}
\end{equation}

The Lie algebroid $L_{(\Sigma,\calf)}$ is given by elements of the form $X- i \, \iota_X \omega$, where $X\in T_\Sigma \otimes \mathbb{C}$, such that $P_\Sigma[\iota_X\omega]=i\,\iota_X\calf$. Thus, we can identify
\begin{equation}
L_{(\Sigma,\calf)} \simeq \{X \in T_\Sigma \otimes \mathbb{C}:\ \ P_\Sigma[\iota_X\omega]=i\,\iota_X\calf\}\ .
\end{equation}
It follows that we have naturally  $T^\perp_\Sigma\subset L_{(\Sigma,\calf)}$ as a subbundle and furthermore  $\calt^{0,1}_\Sigma\simeq L_{(\Sigma,\calf)}/(T^\perp_\Sigma\otimes \mathbb{C})$. Thus we see that we obtain the short exact sequence of bundles
\bea\label{short1}
0\ \rightarrow\  T^\perp_\Sigma\otimes\mathbb{C}\ \rightarrow\  L_{(\Sigma,\calf)}\  \rightarrow\  \calt^{0,1}_\Sigma\ \rightarrow\  0\ .
\eea
To  see the form of the differential $d_{L_{(\Sigma,\calf)}}$, we can locally (and non-canonically) split
\bea\label{coisotropiciso}
 L_{(\Sigma,\calf)}\simeq  (T^\perp_\Sigma\otimes\mathbb{C})\oplus \calt^{0,1}_\Sigma\ ,
\eea
using a local trivialization of the foliation, where we can  uplift a local section $[U]-iP_\Sigma[\iota_U\omega]$ of $\calt^{0,1}_\Sigma$ to a local section  $U-i\iota_U\omega$ of $L_{(\Sigma,\calf)}$ (see \cite{kapustindeform} for a discussion in an explicit coordinate system). In the  same way, using the short exact sequence dual to (\ref{short1}), we can split $L^\star_{(\Sigma,\calf)}$ to obtain the following local form of the Lie algebroid complex $\Lambda^\bullet L^\star_{(\Sigma,\calf)}$
\bea\label{coicomplex}
\Lambda^k L^\star_{(\Sigma,\calf)}\simeq \bigoplus_{p+q=k} \Lambda^pT^{\perp\star}_{\Sigma}\otimes \Lambda^q\calt^{\star 0,1}_\Sigma\ .
\eea
Using the property \eqref{propcourant} of Courant brackets we immediately find the result of \cite{kapustindeform}, namely that (\ref{coisotropiciso})  is a Lie algebroid isomorphism, and thus $d_{L_{(\Sigma,\calf)}}$ acts in the following way on the local form (\ref{coicomplex}) of the complex
\begin{equation}
d_{L_{(\Sigma,\calf)}} \simeq d_{T^\perp_\Sigma} + \bar{\partial}_{ \calt^{0,1}_\Sigma} \, .
\end{equation}
Note that in particular, the massless deformations are given by equivalence classes $[(a,b^{\,0,1})]$ of sections of
\bea
L^\star_{(\Sigma,\calf)}\simeq  \caln^{0,1}_{(\Sigma,\calf)}\simeq  (T^{\perp\star}_{\Sigma}\otimes\mathbb{C})\oplus\calt^{\star 0,1}_\Sigma\ ,
\eea
such that
\bea\label{first}
 \bar{\partial}_{\calt^{0,1}_\Sigma}a + d_{T^\perp_\Sigma}b^{\,0,1} =0\ .
\eea
and
\bea\label{second}
&&d_{T^\perp_\Sigma}a=0\, ,\, a\simeq a+d_{T^\perp_\Sigma}\lambda\ ,\cr
&&\bar{\partial}_{\calt^{0,1}_\Sigma} b^{\,0,1}  =  0\, ,\, b^{\,0,1}\simeq b^{\,0,1}+\bar{\partial}_{\calt^{0,1}_\Sigma} \lambda\ ,
\eea
Since $T^{\perp}_\Sigma\simeq\caln_\Sigma$, $a$ describes the complex combination of the geometric deformation of $\Sigma$ and the component of the  gauge field in $T^{\perp\star}_\Sigma$, while  $b^{\,0,1}$ describes the component of the gauge field deformation in $\calt^{\star 0,1}_\Sigma$.
Thus the first condition (\ref{first}) may be seen as the condition that $\iota_X\calf=0$ for any $X\in T_\Sigma^\perp$ is preserved under the deformation.
On the other hand, note that the first line of (\ref{second}), which controls the directions along the $(d/2-2k)$-dimensional $T^\perp_\Sigma$,
is completely analogous to the result on deformations of special Lagrangian submanifolds, while the second line is reminiscent of the gauge deformations of a B-brane and comes from the requirement that the deformation generated by $b$ preserves the condition that $(\omega|_{\calt_\Sigma})^{-1}\calf|_{\calt_\Sigma}$ squares to $-1$. Note also that the gauge equivalence in the first line of (\ref{second}) tells us that there are gauge deformations of the cycle $\Sigma$ that come from the natural complexification,  induced by the underlying generalized complex geometry, of the (real) world-volume gauge-field transformations along $T_\Sigma^\perp$. These gauge transformations are the coisotropic generalization of the usual Hamiltonian deformations of Lagrangian cycles. They were assumed in \cite{kapustincoisotropic} and we see here how they arise naturally from the generalized complex approach.

We can also immediately find this result
by plugging the deformation generated by $(a,b)$ in \eqref{coisotropiccond}, where ${\rm Im} \, a=i P_{T_\Sigma^\perp}[\iota_X \omega]$  is associated to the deformation of $\Sigma$ generated by $X\in\Gamma[\caln_\Sigma]$ and  $({\rm Re}\,a,b)$ is associated to the deformations of $\calf$. Indeed, we obtain the condition
\begin{equation}
(i P_\Sigma[\omega]+\calf)^k \wedge (da+ db) = 0 \, ,
\end{equation}
which can be shown to be equivalent to (\ref{first}) and (\ref{second}), using the fact that $(iP_\Sigma[\omega]+\calf)^{k}$ descends to a non-vanishing $(2k,0)$ form on $\calt_\Sigma$ .

\subsection{Switching on background fluxes on complex and symplectic manifolds: D-branes on  $SU(3)$ structure backgrounds}
\label{CY3}

In this section we consider non-trivial background fluxes, beginning with flux vacua which preserve an underlying complex or symplectic structure.
A string theory realization of this condition is given by type II flux vacua with  $SU(3)$ structure, that turn out to be symplectic in the IIA case and complex in the IIB case.
These have fluxes turned on, back-reacting on the geometry and generating generically a warp factor.
However, as we have already pointed out, in order to identify the supersymmetric D-brane massless fluctuations,
we can  ignore the complete structure of these backgrounds, focusing only on the complex or symplectic structure of the internal manifolds.

Consider first a symplectic manifold. As we have recalled in subsection \ref{CY1},
the corresponding pure spinor has the form $\psi_s=e^{i\omega}$.
Even if we are considering backgrounds with general fluxes, the only possible modifications involving $\psi_s$ are an overall factor and the $H$-twisting of the integrability condition that now reads $d_H\psi_s=0$.
However, it is easy to see that this integrability condition can only be fulfilled if the overall factor is constant and $H=0$.
So the introduction of fluxes does not modify the analysis of the Lie algebroid structure  at all, and we can directly borrow the results
on A-branes derived in the fluxless background cases studied in sections \ref{fluxlessCY} and \ref{coisotropic}.

Let us now pass to flux complex manifolds and B-branes. The pure spinor has the form $\psi_c=\Omega$,
where $\Omega$ is  a globally defined  $(d/2,0)$ holomorphic form.
However, this time the introduction of general fluxes can have a non-trivial effect on the D-brane Lie algebroids,
since the condition $d_H\psi_c=0$ allows a non-vanishing $H$ field satisfying the restriction $H_{0,3}=H_{3,0}=0$.
In order to compute the relevant algebroid cohomology, we can  parallel the discussion of \cite{kapustindeform} on B-branes in fluxless Calabi-Yau spaces  that we have reviewed in section \ref{Bflux}.
In a similar way, we introduce the same kind of vector-bundle isomorphism leading to
\bea
L_{(\Sigma,\calf)}\simeq T_\Sigma^{0,1} \oplus \caln_\Sigma^{\star 1,0}\ .
\eea
Thus the relevant algebroid graded complex is still given by the global sections of
\bea\label{complex2}
\Lambda^k L_{(\Sigma,\calf)}^\star\simeq \Lambda^k\caln_{(\Sigma,\calf)}^{0,1}\simeq\bigoplus_{p+q=k}\Lambda^pT_\Sigma^{\star 0,1}\otimes \Lambda^q\caln_\Sigma^{1,0}\ .
\eea
The computation of the Lie algebroid differential $d_{L_{(\Sigma,\calf)}}$ proceeds as in the $H=0$ case,
with the only differences that we must use the $H$-twisted Courant bracket (\ref{Hcourant})  and take into account that $d\calf=P_{\Sigma}[H]$.
The result is that
\bea
d_{L_{(\Sigma,\calf)}}\simeq\bar\partial +\delta^{H}(\Sigma,\calf)\ ,
\eea
with
\bea\label{twisteddelta}
\delta^{H}(\Sigma,\calf)=\delta(\Sigma,\calf)+ P_\Sigma[H^{1,2}\llcorner]\ ,
\eea
where $\delta(\Sigma,\calf)$ is as defined in (\ref{beta}) and  $\llcorner$  contracts the  $(1,0)$ index in $H^{1,2}$  with the indices in $\Lambda^q\caln_\Sigma^{1,0}$ of the sections of (\ref{complex2}).\footnote{Let us stress that the action of $\llcorner$ in $P_\Sigma[H^{1,2}\llcorner]$, as the operator $\delta(\Sigma,\calf)$ itself,
is well-defined once we have fixed a (non-canonical) choice for the uplifting of the holomorphic sections of $\caln_{\Sigma}^{1,0}$
to smooth sections of $T^{1,0}_M|_\Sigma$, as given by the local sections $p$ introduced in (\ref{exact}).
It is however important to note that, thanks  to the modified Bianchi identity $d\calf=P_\Sigma[H]$,
the operator $\delta^{H}(\Sigma,\calf)$, seen as a section of $\Lambda^2 T^{\star 0,1}_\Sigma\otimes \caln_\Sigma^{\star 1,0}$,
is $\bar\partial$-closed, and that the corresponding cohomology element of $H^{0,2}_{\bar{\partial}}( \caln_\Sigma^{\star 1,0})$
is `canonical' in the sense that it does not depend on the choice of the $0$-chain $p$.} Thus, we repeat the steps of the case with $H=0$, arriving at the following massless spectrum:
\bea\label{fluxfluctuationsH}
H^1(L_{(\Sigma,\calf)})= H^{0,1}_{\bar\partial}(\Sigma)\oplus  H^0_{d_{L_{(\Sigma,\calf)}}}(\caln^{1,0}_\Sigma)\ .
\eea
where \bea\label{gmasslessH}
H^0_{d_{L_{(\Sigma,\calf)}}}(\caln^{1,0}_\Sigma)={\rm ker}[ \delta^H(\Sigma,\calf): H^0_{\bar\partial}(\caln^{1,0}_\Sigma)\rightarrow H^{0,2}_{\bar\partial}(\Sigma)]
\eea

Again, as in the case $H=0$, $H^{0,1}_{\bar\partial}(\Sigma)$ in (\ref{fluxfluctuationsH}) corresponds to the unlifted world-volume gauge field deformations.
Note however that, differently from the case $H=0$, now
there is still a generically  non-vanishing moduli-lifting effect even if $T_M^{1,0}$ holomorphically splits into $T_\Sigma^{1,0}\oplus \caln_\Sigma^{1,0}$. This effect is completely due to the non-vanishing $H$ field, that can give mass to all the otherwise massless geometrical modes.
This fact was already noticed for warped Calabi-Yau backgrounds in \cite{marchesano1},
and further clarified for more general K\"ahler  $SU(3)$ structure vacua in \cite{branesuppot} using the D7-brane superpotential
that still has the form (\ref{hol4}). Indeed, the superpotential (\ref{hol4}) gives automatically the condition (\ref{gmasslessH}) for the massless geometrical modes, also in the more general case that   $T_M^{1,0}|_\Sigma$ does not holomorphically splits into $T_\Sigma^{1,0}\oplus \caln_\Sigma^{1,0}$, since the corresponding holomorphic mass matrix is given by
\bea
m_{ij}^{\rm (hol)}&=&\partial_i\partial_j\calw=\int_{\Sigma_4} P_{\Sigma_4}[\iota_{X_i}\Omega]\wedge\big(P_\Sigma[\iota_{X_j}H]  -[(\bar\partial X_j)\circ \calf]\big)=\cr
&=&\int_{\Sigma_4} P_{\Sigma_4}[\iota_{X_i}\Omega]\wedge [\delta^H(\Sigma,\calf)\cdot X_j] \ ,
\eea
where the $[X_i]$ form a base of $H^0_{\bar{\partial}}(\caln_\Sigma^{1,0})$.


\subsection{D3-brane moduli lifting in a background with an `honest' generalized complex structure.}
\label{CY5}

In this section we would like to analyse a case in which the background has an `honest' generalized complex structure, in the sense that it is neither complex nor symplectic.

For definiteness, we will focus on the six-dimensional case with a generalized complex structure of odd type, associated to a $d_H$-closed pure spinor of the general form
 \bea\label{oddpspinor}
 \psi=\psi_{(1)}+\psi_{(3)}+\psi_{(5)}\ .
 \eea
If $\psi_{(1)}$ is nowhere zero, this situation would occur in a type IIB background with global $SU(2)$ structure. However, in this case we
allow $\psi_{(1)}$ to become zero at certain points, which we call supersymmetric points for reasons that will become clear in a moment.

We would like to apply our general results to identify the massless deformations of a D3-brane located at a supersymmetric point $y_0\in M$.
This kind of situation, even if it involves the simplest D-brane one can consider, will nevertheless show interesting features directly due to the `non-trivial' underlying generalized complex structure.
For example, one key difference with respect to the cases discussed in the previous subsections is that we are going to compute the first cohomology group
of a Lie algebroid on a {\em point}.
Nevertheless, from the analysis of \cite{branesuppot} one can immediately see that, differently from the point-like B-branes on complex manifolds, the Lie algebroid differential must be non-trivial, thus leading to a first cohomology group which differs from the tangent space $T_M|_{y_0}$ itself.

It is convenient to address the problem by first looking at the D$3$-brane superpotential. From the general formula  (\ref{supot}),
we find that the D$3$-brane superpotential $\calw_{\rm D3}$ is simply related to the one-form  $\psi_{(1)}$ appearing in (\ref{oddpspinor}) by the formula
\bea\label{d3sup}
\psi_{(1)}=d\calw_{\rm D3}\ .
\eea
It is then  clear that the D$3$-brane superpotential is non-trivial if and only if $\psi_{(1)}$ does not identically vanish\footnote{For explicit examples in the gauge/gravity correspondence context where this happens  see  \cite{mpz} , which studies $SU(2)$ structure backgrounds dual to deformations of $\caln=4$ Super Yang-Mills theory.}.
If it does vanish everywhere, $\psi$ can be locally put in the form of a $B$-transformed holomorphic $(3,0)$-form,
thus  defining an ordinary complex structure.
On the other hand, from (\ref{d3sup}) one can also immediately see that a D$3$-brane can only be supersymmetric at $y_0\in M$ if $\psi_{(1)|y_0}=0$.
This means that the D$3$-brane must be located at a point where the type of the generalized complex structure jumps from one to three\footnote{We recall that the type of a generalized complex structure is an upper semicontinuous function on the manifold, meaning that each point has a neighbourhood where it does not decrease \cite{gualtieri}.}.
As we will see, this interesting feature will characterize our calculation of the Lie algebroid cohomology.

At $y_0$, up to a $B$-transformation, the pure spinor $\psi$ reduces to $(3,0)$-form
\bea
\psi_{|y_0}=\Omega_{|y_0}
\eea
associated to a complex structure $J_{|y_0}$ on $T_M|_{y_0}$.
Since we are interested in small fluctuations around $y_0$, we can restrict to a small neighbourhood $U$ of $y_0$. If $U$ is small enough, we can extend $J_{|y_0}$ to an {\em integrable} complex structure $J$ on $U$ and  $\Omega_{|y_0}$ to the corresponding holomorphic $(3,0)$-form $\Omega$,
and consider $\psi$ on $U$ as some small deformation  of $\Omega$.

The deformation theory of generalized complex structures has been studied in detail in \cite{gualtieri}, and in particular an infinitesimal deformation of an ordinary complex structure can be of three different kinds,
corresponding respectively to an ordinary deformation of the complex structure, a $B$-field transformation, and a $\beta$-deformation.
While the first two deformations do not change the type of the generalized complex structure, the third does.
So, up to a possible $B$-transformation, without loosing generality we can consider the pure spinor $\psi$ on $U$ as given by a  $\beta$-deformation of $\Omega$, where $\beta$ is a holomorphic section of $\Lambda^2T^{1,0}_U$ vanishing at $y_0$ and defining a Poisson structure:
\bea
\bar\partial\beta=0\, ,\qquad [\beta,\beta]=0\ .
\eea

Now, at $y_0$ the maximal isotropic subspace associated to $\psi$ is simply given by
\bea
L|_{y_0}=T_M^{0,1}|_{y_0}\oplus T_M^{\star 1,0}|_{y_0}\ ,
\eea
while the generalized tangent bundle of the cycle $\{y_0\}$ is  given by
\bea
T_{\{y_0\}}=T^\star_M|_{y_0}\ .
\eea
Thus the Lie algebroid associated to the (generalized) zero-cycle $\{y_0\}$ is simply given by
\bea
L_{\{y_0\}}=T_M^{\star 1,0}|_{y_0}\ .
\eea
To describe the Lie brackets between vectors  of this algebroid,  we need to extend them outside $y_0$ to local sections of $L$. As we explained above, restricting to a small enough neighbourhood $U$,  $L$ can be approximated, up to a possible $B$-transformation, by a $\beta$-deformation of $T_U^{0,1}\oplus T_U^{\star 1,0}$.  In particular we can consistently extend elements $\eta\in T_M^{\star 1,0}|_{y_0}$ to $\eta-\iota_\beta\eta$ on $U$, the $B$-transformation having no effect on them.

For a base of these extended elements of the Lie algebroid, given in
terms of complex coordinates on $U$ as $e^i=dz^i+\beta^{ij}\partial_j$, the Courant brackets are given by
\bea
[e^i,e^j]=(\partial_l\beta^{ij})_{|{y_0}} e^l\ .
\eea
The graded differential complex is given by $\bigoplus_k \Lambda^k T_M^{1,0}|_{y_0}$, and
the algebroid differential
\bea
d_{L_{\{y_0\}}}: \Lambda^k T_M^{1,0}|_{y_0}\rightarrow \Lambda^{k+1} T_M^{1,0}|_{y_0}
\eea
acts on elements
\bea
\alpha=\frac{1}{k!}\alpha^{i_1\ldots i_k}\partial_{i_1}\wedge\ldots\partial_{i_k}\big|_{y_0}\in \Lambda^k T_M^{1,0}|_{y_0}\ ,
\eea
as follows
\bea
d_{L_{\{y_0\}}}\alpha=- \partial\beta\circ \alpha\equiv -\frac{1}{2 (k-1)!}\,\partial_l\beta^{i_1i_2}\alpha^{li_3\ldots i_{k+1}}\partial_{i_1}\wedge\ldots\wedge\partial_{i_{k+1}} \big|_{y_0} \ .
\eea

We can thus conclude that the cohomology group $H^1(L_{\{y_0\}})$ giving the massless fluctuations of the D$3$-brane is given by $(1,0)$ vectors $X$ at $y_0$, such that
\bea\label{d3massless}
\partial\beta\circ X\equiv \frac{1}{2}\,\partial_l\beta^{ij} X^l\partial_{i}\wedge\partial_{j} \big|_{y_0}=0\ .
\eea

Physically, the same answer for the D$3$-brane massless fluctuations can be directly extracted from the superpotential  (\ref{d3sup}). It is enough to observe that on $U$ we can approximate
\bea
\psi_{(1)}=\iota_\beta\Omega\ .
\eea
Using the fact that $\beta_{|y_0}=0$, the holomorphic mass-matrix is given by
\bea
m_{ij}=(\partial_i\partial_j\calw_{\rm D3})_{|y_0}=\partial_i\psi_j=\frac{1}{2} \, [(\partial_i\beta^{kl})\Omega_{klj}]_{|y_0}\ ,
\eea
which is symmetric  since $\partial(\iota_\beta\Omega)=0$. We see that the zero-eigenvectors of this mass matrix are given by the vectors $X\in T^{1,0}_M|_{y_0}$ such that  (\ref{d3massless}) is satisfied, thus reproducing the result obtained from the Lie algebroid cohomology.


\section{Conclusions}

We have shown that the deformations of supersymmetric D-branes that preserve supersymmetry in a generalized complex vacuum are described by the  first cohomology group $H^1(L_{(\Sigma,\calf)})$, associated to the D-brane Lie algebroid  $L_{(\Sigma,\calf)}$.  This Lie algebroid has a natural complex structure induced by the underlying generalized complex structure $\calj$.
Thus $H^1(L_{(\Sigma,\calf)})$ is also naturally a complex vector space. This makes sense as in the four-dimensional low-energy description the massless fluctuations should organize into the complex scalars of chiral multiplets.

We conclude with a few remarks and avenues for further research.
First of all, we have not addressed the issue of higher order obstructions.
We know that our formalism includes special Lagrangian and complex cycles as particular subcases in Calabi-Yau manifolds.
The former are known to be unobstructed, while for the latter the possible obstruction is encoded in a non-trivial element of $H^1(\caln_\Sigma^{1,0})$.
Thus, generically we expect the massless deformations associated to $H^1(L_{(\Sigma,\calf)})$ to be obstructed at higher order. It would be very interesting to see if it is possible to associate  possible higher order obstructions to the existence of some non-trivial cohomology group connected to the D-brane, as it happens for complex cycles. As discussed in section \ref{4d}, the geometrical superpotential  (\ref{supot}) provides an alternative point of view on this problem. This superpotential is the direct generalized complex analog of Witten's holomorphic Chern-Simons action \cite{wittenCS}  and can in principle generate the complete (classical) effective superpotential for the four-dimensional massless fields associated  to    $H^1(L_{(\Sigma,\calf)})$. This effective superpotential should contain all the information about higher order obstructions. However, as for the  holomorphic Chern-Simons action, a straightforward computation of the effective superpotential from (\ref{supot}) is not easy,
as it seems to involve the complete $SU(3)\times SU(3)$ structure of the background.  A more efficient and natural method should rely only  on the underlying generalized complex structure,
like for example the method proposed for B-branes in \cite{ak}, which depends only on the algebraic geometry of the underlying Calabi-Yau.

To understand some characteristic features of our generalized setting that are new with respect to the Calabi-Yau subcase,
we have explicitly considered  a D3-brane in a type-changing generalized complex structure. This example is interesting for two reasons. First of all, it tests the limits of our formalism and shows that to properly define the Lie brackets on $L_{(\Sigma,\calf)}$ we need to extend the sections of the Lie algebroid outside the D3-brane. The result does not depend on the choice of extension, but it {\em does} depend on the form of the generalized complex structure in a neighbourhood of the D3-brane. Secondly, it considers a case in which some of the moduli are lifted and the D3-brane is not free to move anywhere in the internal manifold. This kind of effect is induced by the underlying `honest' generalized complex structure (of odd type) and could potentially be interesting for phenomenological models as it provides a mechanism for ``freezing'' the position of the D3-brane.\footnote{This effect is already evident from the D$3$-brane superpotential, which was first derived in \cite{branesuppot},
and has for example been applied in \cite{mpz} in the context of  the gauge/gravity correspondence.}

Other peculiar effects due to the generalized nature of the underlying geometry should arise also in more complicated setting, involving higher-dimensional D-branes. For example, it would be interesting to make a systematic analysis of generalized calibrations in the backgrounds of \cite{grananil}. Furthermore other non-trivial examples would be D-branes in group manifolds \cite{lindstromWZW}.

Another feature of our analysis is that it is completely symmetric under the simultaneous exchange of type IIA and IIB and of the even and odd pure spinors characterizing the $SU(3)\times SU(3)$ structure. Our results should thus be helpful to gain a better understanding of the extension of mirror symmetry to generalized complex flux vacua, in the spirit of \cite{syz}.\footnote{See \cite{minasian1,minasian2} for related work on D-branes in generalized geometries.}  Also, there is a natural and interesting interplay with non-geometrical backgrounds \cite{hull}, which deserves further investigation. For this research a better understanding of the properties of D-branes on generalized complex manifolds can provide a key tool, as they  are the  natural probe objects to investigate these issues (see for example \cite{grange} for a discussion in this direction).

A very challenging problem is to try to extend the analysis to coinciding D-branes for which the gauge bundle becomes non-abelian. It is known  that, at least for a nine-dimensional B-brane, the F-flatness condition stays the same while the D-brane action
and the D-flatness condition become horrendously complicated, see \cite{koerberthesis} for a review and further references. Moreover, the coordinates describing the position of coinciding D-branes become matrix-valued,
and, as is well-known, complicate an intrinsic geometrical formulation of problem \cite{wittenmatrix,douglasmatrix,deboermatrix,raamsdonkmatrix}.
Again, a possible simplification may arise by focusing on the holomorphic sector of the theory,
which should still depend only on the integrable generalized complex structure of the underlying flux background.

Finally, in this paper we have considered D-branes on fixed flux backgrounds,
whose possible closed string moduli have been frozen.
On the other hand, the effective description of the closed string sector in the same setting has been started in \cite{grana,grimm}.
The natural subsequent step would be to glue together the two approaches, and obtain a description of the complete bulk-plus-branes system.

\vspace{1cm}

\section*{Acknowledgements}
L.~M.\ thanks  A.~Tomasiello  for useful discussions.  P.~K.\ wishes to thank the University of British Columbia, which was his home institution
during the bulk of this work. L.~M.\ is supported in part by the Federal Office for Scientific, Technical and Cultural Affairs through the ``Interuniversity Attraction Poles Programme -- Belgian Science Policy" P5/27 and by the European Community's Human Potential Programme under contract
MRTN-CT-2004-005104 `Constituents, fundamental forces and symmetries of the universe'.

\vspace{1cm}

\begin{appendix}

\section{D-flatness and gauge fixing}
\label{mm}

In \cite{branesuppot} it was shown how the D-flatness condition (\ref{dflatness}) can be seen as the vanishing of the moment map associated to the world-volume gauge transformations, and thus provides a gauge fixing condition for the imaginary extension of the world-volume gauge transformations. In this appendix we recall the definition of such a moment map and show how it can be used to see that any infinitesimal deformation of a generalized calibrated cycle can always be `corrected' with an additional  imaginary gauge transformation, so that the D-flatness condition is still satisfied.

The symplectic structure $\Xi$ necessary to obtain such a moment map characterization
of the D-flatness condition was introduced for an arbitrary generalized cycle by using the background $SU(3)\times SU(3)$ structure
to choose a particular representative for each vector of $\caln_{(\Sigma,\calf)}$. However, if we restrict to a generalized calibrated cycle $(\Sigma,\calf)$, it is possible
to write down the following canonical form
\bea
\Xi([\mathbb{X}],[\mathbb{Y}])|_{(\Sigma,\calf)}=\int_\Sigma P_\Sigma[e^{2A-\Phi}\mathbb{X}\cdot\mathbb{Y}\cdot{\rm Im}\,\hat\Psi_1]\wedge e^\calf\ ,
\eea
and one can check that the density
\bea\label{mmap}
m_{(\Sigma,\calf)}=P_\Sigma[e^{2A-\Phi}{\rm Im}\,\hat\Psi_1]\wedge e^\calf|_{\rm top}\ ,
\eea
indeed provides the moment map associated to the world-volume gauge transformations, i.e., for any $\lambda\in\Gamma(\Sigma,\mathbb{R})$ and  $[\mathbb{Y}]\in \Gamma(\caln_{(\Sigma,\calf)})$ we have:
\bea
\iota_{\mathbb{Y}}d(m_{(\Sigma,\calf)}(\lambda))=\Xi(\mathbb{X}_{\lambda},\mathbb{Y})\ .
\eea

One can also introduce the following metric structure on $\caln_{(\Sigma,\calf)}$
\bea\label{metric}
G([\mathbb{X}],[\mathbb{Y}])\equiv \Xi(\calj[\mathbb{X}],[\mathbb{Y}])\ .
\eea
If we impose the gauge-fixing that puts $\caln_{(\Sigma,\calf)} \simeq C_+|_\Sigma$ using the $SU(3)\times SU(3)$ structure, where $C_+\subset T_M\oplus T^\star_M$ is the sub-bundle of the  generalized vectors of the form $\mathbb{X}^+=(X,g\cdot X)$  (see \cite{gualtieri,branesuppot} for more details),  the above metric reduces to
\bea\label{metric2}
G(\mathbb{X}^+,\mathbb{Y}^+)|_{(\Sigma,\calf)}=\int_\Sigma P_\Sigma[\cali(\mathbb{X}^+,\mathbb{Y}^+)e^{2A-\Phi}{\rm Re}\,\hat\Psi_1]\wedge e^\calf\ ,
\eea
which is the metric on $\caln_{(\Sigma,\calf)}$ in the `gauge fixed' form written down in \cite{branesuppot},
from which we explicitly see that it is positive definite since restricting to $C_+$ makes $\cali$ positive definite
and $P_\Sigma[e^{-\Phi}{\rm Re}\,\hat\Psi_1]\wedge e^\calf$
is the volume form on a generalized calibration. More generally, we can still use the metric (\ref{metric})
which is non-degenerate and positive definite for non-calibrated generalized cycles which are close enough to a calibrated one.

By standard arguments, one sees that the D-flatness condition provides a slice of the imaginary gauge transformations.
Indeed, an imaginary gauge transformation generated by $\lambda\in\Gamma(\Sigma,\mathbb{R})$ takes us away from the D-flatness condition $m_{(\Sigma,\calf)}=0$
since
\bea
\iota_{\calj\mathbb{X}_\lambda}d( m_{(\Sigma,\calf)}(f))=\Xi(\calj\mathbb{X}_\lambda,\mathbb{X}_f)=G(\mathbb{X}_\lambda,\mathbb{X}_f)
\eea
cannot be zero for all $f$. It suffices to take $f=\lambda$ to see this.
This can also be shown directly at the infinitesimal level starting from the condition (\ref{dvar}), which implies that for any $f\in\Gamma(\Sigma,\mathbb{R})$ we should have
\bea\label{dvar2}
0&=&\int fP_\Sigma[d_H(e^{2A-\Phi}\mathbb{X}\cdot{\rm Im}\,\hat\Psi_1)]\wedge e^\calf|_{\rm top}=\cr
&=& -\int df\wedge P_\Sigma[e^{2A-\Phi} (\mathbb{X}\cdot{\rm Im}\,\hat\Psi_1)]\wedge e^\calf|_{\rm top}=\Xi(\mathbb{X},\mathbb{X}_f)\ .
\eea
If we now take the imaginary gauge transformation $[\mathbb{X}]=\calj[\mathbb{X_\lambda}]\equiv \calj[d\lambda]$ we arrive at
\bea
\Xi(\calj\mathbb{X}_\lambda,\mathbb{X}_f)=G(\mathbb{X}_\lambda,\mathbb{X}_f)\ ,
\eea
which clearly does not vanish for all $f$. Again, it is non-zero at least for $f=\lambda$.

Furthermore this formalism allows us to see that it is always possible to find an element in the imaginary gauge orbit of an $[\mathbb{X}]\in\caln_{(\Sigma,\calf)}$ satisfying the condition (\ref{defcond}) which also satisfies the D-flatness condition. Indeed we can square $m_{(\Sigma,\calf)}$ using the inverse of the metric defined on the dual space (see \cite{branesuppot} for an explicit expression derived from the Dirac-Born-Infeld action)
and then $m_{(\Sigma,\calf)}=0$ if and only if $||m||^2_{(\Sigma,\calf)}=0$.
Now, we see that
\begin{equation}
\call_{[\mathbb{X}]}(-||m||^2)|_{(\Sigma,\calf)}=-\call_{[\mathbb{X}]}(m(m^\star))|_{(\Sigma,\calf)}=-\Xi(\mathbb{X}_{m^\star},\mathbb{X})|_{(\Sigma,\calf)}=G(\mathbb{X},\calj \mathbb{X}_{m^\star})|_{(\Sigma,\calf)}\ .
\end{equation}
This shows that the gradient flow of $-||m||^2$ is along the imaginary gauge orbits and in particular, for any $[\mathbb{X}]$ we can take a $\mathbb{Y}$ of the form
\bea
\mathbb{Y}=-\frac{G(\mathbb{X},\calj\mathbb{X}_{m^\star})}{G(\calj\mathbb{X}_{m^\star},\calj\mathbb{X}_{m^\star})}\calj\mathbb{X}_{m^\star}\ ,
\eea
generating an imaginary gauge transformation, such that
\bea
\call_{[\mathbb{X}+\mathbb{Y}]}(-||m||^2)|_{(\Sigma,\calf)}=0\ .
\eea


\section{The $d^\dagger_{L_{(\Sigma,\calf)}}$ operator}
\label{hodge}

In appendix \ref{mm} we introduced the metric (\ref{metric}) on the space of sections of $\caln_{(\Sigma,\calf)}$, and thus of $T^\star_{(\Sigma,\calf)}$. We can now extend this metric to the space of sections of $\Lambda^k T^\star_{(\Sigma,\calf)}$ in the following way. Use the identifications $T^\star_{(\Sigma,\calf)}\simeq \caln_{(\Sigma,\calf)}\simeq C_+|_\Sigma\simeq T_M|_\Sigma$ (for more details about the second identification see \cite{branesuppot}) to expand a generic section $\alpha$ of $\Lambda^k T^\star_{(\Sigma,\calf)}$ as follows
\bea
\alpha=\frac{1}{k!}\alpha^{m_1\ldots m_k}\partial_{m_1}{}_{|\Sigma}\wedge\ldots\wedge \partial_{m_k}{}_{|\Sigma}\ .
\eea
Now we can define the inner product between two sections $\alpha,\beta$ of $\Lambda^k T^\star_{(\Sigma,\calf)}$ in the following way
\bea
G(\alpha,\beta)=\frac{1}{k!}\int_M\alpha^{m_1\ldots m_k}\beta_{m_1\ldots m_k} \langle e^{2kA- \phi} {\rm Re} \,  \hat{\Psi}_1,j_{(\Sigma,\calf)}\rangle \, ,
\label{metricforms}
\eea
where the indices of $\beta$ are lowered using the metric $g_{mn}$ defined on $M$ by the $SU(3)\times SU(3)$ structure.  Note the dependence in (\ref{metricforms}) of the factor $e^{2kA}$ on the dimension $k$ of the forms, and the fact that it obviously reduces to (\ref{metric2}) in the case of one-forms.

This metric can also be used as a positive-definite metric on sections of $\Lambda^k L^\star_{(\Sigma,\calf)}$ by simply adding a complex conjugation on the second form.
Thus, we can define the Lie algebroid codifferential
\bea
d^\dagger_{L_{(\Sigma,\calf)}}: \Gamma(\Lambda^k L^\star_{(\Sigma,\calf)})\rightarrow  \Gamma(\Lambda^{k-1} L^\star_{(\Sigma,\calf)})
\eea
in the usual way:
\bea
G(d^\dagger_{L_{(\Sigma,\calf)}}\alpha,\overline\beta)=G(\alpha,\overline{d_{L_{(\Sigma,\calf)}}\beta})\quad \text{for\ any}\quad  \beta\in\Gamma(\Lambda^{k-1} L^\star_{(\Sigma,\calf)})\ .
\eea

As an example, it is easy to see that the complexified D-flatness condition (\ref{complexcond}) can be expressed in terms of the codifferential $d^\dagger_{L_{(\Sigma,\calf)}}$. Indeed, it follows from that condition that we must have for any $f\in\Gamma(\Sigma,\mathbb{C})$
\bea
0&=& - i\int_M \langle \bar f {\partial}_H(e^{2A-\Phi}{\mathbb{X}^{0,1}}\cdot{\rm Im}\,\hat\Psi_1),j_{(\Sigma,\calf)}\rangle=i\int_M \langle
e^{2A-\Phi}\,\overline{d_{L_{(\Sigma,\calf)}} f}\cdot {\mathbb{X}^{0,1}}\cdot{\rm Im}\,\hat\Psi_1,j_{(\Sigma,\calf)}\rangle\cr
&=&G([\mathbb{X}^{0,1}],\overline{d_{L_{(\Sigma,\calf)}}f})=G(d_{L_{(\Sigma,\calf)}}^\dagger [\mathbb{X}^{0,1}],\bar f)\ ,
\eea
and thus the condition (\ref{complexcond}) can be written in the form
\bea
d_{L_{(\Sigma,\calf)}}^\dagger [\mathbb{X}^{0,1}]=0\ .
\eea

We can also show that the $d_{L_{(\Sigma,\calf)}}$-complex is elliptic, so that its cohomology classes on a compact cycle $\Sigma$ are finite-dimensional.
The anchor map $\pi:L_{(\Sigma,\calf)}\rightarrow T_\Sigma\otimes \mathbb{C}$ is the obvious projection to $T_\Sigma\otimes \mathbb{C}$ and the principal symbol (see e.g.\ \cite{egh} for a review on elliptic complexes) of $d_{L_{(\Sigma,\calf)}}$,
\begin{equation}
s(d_{L_{(\Sigma,\calf)}}):T^\star_\Sigma \otimes \Lambda^k L^\star_{(\Sigma,\calf)}
\rightarrow \Lambda^{k+1} L^\star_{(\Sigma,\calf)}  \, ,
\end{equation}
is given by
\begin{equation}
s_\xi(d_{L_{(\Sigma,\calf)}}) = \pi^\star(\xi) \wedge \cdot ,
\end{equation}
with $\xi$ a one-form of $T^\star_\Sigma$. Since $\xi$ is real, if $\xi\neq 0$ then one can easily see that $\pi^\star(\xi)\equiv \mathbb{X}$ is a non-zero element of $L^\star_{(\Sigma,\calf)}\simeq \caln^{0,1}_{(\Sigma,\calf)}$. It follows that $s_\xi(d_{L_{(\Sigma,\calf)}})$ clearly defines an exact complex and thus the differential complex defined by $d_{L_{(\Sigma,\calf)}}$ is elliptic. Equivalently, since the symbol of the
Laplacian \eqref{laplacian}, given by $s_\xi(\Delta_{L_{(\Sigma,\calf)}}) = -|\mathbb{X}|^2$ (where one uses the metric $g_{mn}$ after the identification $\caln_{(\Sigma,\calf)}\simeq C_+|_\Sigma\simeq T_M|_\Sigma$), is invertible, the complex is elliptic.


\section{Masses of the fluctuations}
\label{masses}

In this section we calculate the second variation of the D-brane potential around a calibrated/supersymmetric configuration. This will give us the
classical masses of the fluctuations. We find that, as expected, the massless deformations must satisfy the conditions \eqref{adjoint},
i.e. they must keep the generalized cycle  calibrated. For complex submanifolds without fluxes the result is known as Simons' formula \cite{simons}. It was generalized to ordinary calibrations in \cite{mclean} and we generalize it here to our generalized
calibrations. We perform the calculation for 6 dimensions as appropriate for string theory compactified to 4-dimensional Minkowski space, but the calculation can be extended to different dimensions, at least in the case of zero RR-fluxes.

The four-dimensional potential for a D-brane wrapping an $n$-dimensional generalized cycle $(\Sigma,\calf)$ is given by
\begin{equation}\label{potential}
{\calv}_{(\Sigma,\calf)} = \int_\Sigma d^n \sigma \, e^{4A-\Phi} \sqrt{\text{det}(P_\Sigma[g]+\calf)} - \int e^{4A} \tilde{C} \wedge e^\calf \, .
\end{equation}
To consider its quadratic expansion around a supersymmetric configuration, it is useful to use a notation analogous to the one used for standard $\caln=1$ four-dimensional theories. So we redefine the moment map (\ref{mmap}) as a D-term density
\bea
\cald_{(\Sigma,\calf)}\equiv m_{(\Sigma,\calf)}\ .
\eea
$\cald_{(\Sigma,\calf)}$ can be seen as an element of the space dual to the space of functions on $\Sigma$, where the pairing is given by the standard integration. In the same way, the one-form $\theta=d\calw$ on the configuration space, introduced in section \ref{4d}, can be seen as an element of the space dual to the space of sections of the generalized normal bundle  $\caln_{(\Sigma,\calf)}$. In particular, we can square $\cald$ and $\theta$ using the inverse of the metric (\ref{metricforms}) for functions on $\Sigma$ and sections of $\caln_{(\Sigma,\calf)}$.

Consider now a small expansion around a supersymmetric configuration $(\Sigma_0,\calf_0)$. It was shown in \cite{branesuppot} that the potential (\ref{potential}) takes the following form:\footnote{Here we are using the conventions of the present paper, which amounts to a factor of $1/2$ in front of the superpotential
with respect to \cite{branesuppot}. Also, the metric for the functions on $\Sigma$ was denoted by $k$ in there instead.}
\bea\label{potvar}
{\calv}_{(\Sigma,\calf)} \simeq {\calv}_{(\Sigma_0,\calf_0)}+\frac12 G^{-1}(\cald,\cald)+\frac14 G^{-1}(\theta,\overline\theta)\ ,
\eea
where ${\calv}_{(\Sigma_0,\calf_0)}$ is a constant corresponding to the minimal energy.
More precisely, we can consider a small neighbourhood of $(\Sigma_0,\calf_0)$ as parametrized by  sections $[\mathbb{X}]$ of $\caln_{(\Sigma_0,\calf_0)}$. Then we have
\bea\label{cder}
\cald_{(\Sigma,\calf)}\simeq \nabla_{[\mathbb{X}]}\cald|_{(\Sigma_0,\calf_0)}\quad,\quad
\theta_{(\Sigma,\calf)}\simeq \nabla_{[\mathbb{X}]}\theta|_{(\Sigma_0,\calf_0)}\ ,
\eea
where the covariant derivatives do not actually depend on the choice of a connection since $\cald|_{(\Sigma_0,\calf_0)}=0$ and $\theta|_{(\Sigma_0,\calf_0)}=0$. The explicit expressions for $\nabla_{[\mathbb{X}]}\cald|_{(\Sigma_0,\calf_0)}$ and $\nabla_{[\mathbb{X}]}\theta|_{(\Sigma_0,\calf_0)}$  are given by their pairing with any function $f$ on $\Sigma_0$ and any section $[\mathbb{Y}]$ of $\caln_{(\Sigma_0,\calf_0)}$ respectively. In particular, we obtain
\bea\label{inf1}
\nabla_{[\mathbb{X}]}\cald|_{(\Sigma_0,\calf_0)}(f)&=& \int_{\Sigma_0} f P_{\Sigma_0}[2 \, {\rm Re}\,\{ \partial_H(e^{2A-\Phi}\mathbb{X}^{0,1}\cdot {\rm Im}\,\hat\Psi_1)\}]\wedge e^{\calf_0}=\cr
&=& 2 \, G(f,{\rm Re}\,(d^\dagger_{L_{(\Sigma_0,\calf_0)}}[\mathbb{X}^{0,1}]))\ ,
\eea
and
\bea\label{inf2}
\nabla_{[\mathbb{X}]}\theta|_{(\Sigma_0,\calf_0)}([\mathbb{Y}])&=&\int_{\Sigma_0}P_{\Sigma_0}[e^{3A-\Phi}\mathbb{Y}^{0,1}\cdot\bar\partial_H(\mathbb{X}^{0,1}\cdot \hat\Psi_2)]\wedge e^{\calf_0}\ .
\eea
Plugging in (\ref{cder}) and (\ref{inf1}) into (\ref{potvar}), we get
\bea
\label{potvar2}
{\calv}_{(\Sigma,\calf)} &\simeq& {\calv}_{(\Sigma_0,\calf_0)}+\frac14 G^{-1}(\nabla_{[\mathbb{X}]}\theta,\overline{\nabla_{[\mathbb{X}]}\theta})|_{(\Sigma_0,\calf_0)}+\cr
&&+2 \,  G({\rm Re}\,(d^\dagger_{L_{(\Sigma_0,\calf_0)}}[\mathbb{X}^{0,1}]), {\rm Re}\,(d^\dagger_{L_{(\Sigma_0,\calf_0)}}[\mathbb{X}^{0,1}]))|_{(\Sigma_0,\calf_0)}\ .
\eea

From (\ref{inf2}) it is clear that $\nabla_{[\mathbb{X}]}\theta|_{(\Sigma_0,\calf_0)}$ vanishes as a one-form on the configuration space if and only if $d_{L_{(\Sigma_0,\calf_0)}}[\mathbb{X}^{0,1}]=0$. Imposing our standard gauge-fixing of the real gauge transformations of the form
\bea
{\rm Im}\,(d^\dagger_{L_{(\Sigma_0,\calf_0)}}[\mathbb{X}^{0,1}])=0\ ,
\eea
we can see (\ref{potvar2}) as giving the mass of the gauge-inequivalent fluctuations. In particular, (\ref{potvar2}) proves our claim, namely   that $[\mathbb{X}]$ describes a massless fluctuation if and only if
\bea
d_{L_{(\Sigma_0,\calf_0)}}[\mathbb{X}^{0,1}]=0\quad,\quad d^\dagger_{L_{(\Sigma_0,\calf_0)}}[\mathbb{X}^{0,1}]=0\ .
\eea

\section{Comments on the FI-term and stability}
\label{stability}

We relegate to this appendix  a short comment on the problem of stability that we have anticipated in section \ref{4d}.
In our framework we have single D-branes on an $\caln=1$ background and we expect to have a purely $\caln=1$ description. In particular, if we are in the geometrical regime where we have no change in the spectrum due to stringy effects, the massless four-dimensional $U(1)$ gauge symmetry has no charged matter and then a necessary condition for having a supersymmetric vacuum is that the Fayet-Iliopoulos term $\xi$ must vanish. An explicit expression for the Fayet-Iliopoulos term was given in \cite{branesuppot} and depends on the non-integrable pure spinor alone\footnote{Again, we neglect the correct dimensionful normalizations (see \cite{branesuppot} for the precise expressions).}:
\bea
\xi=\int_\Sigma P_\Sigma[e^{2A-\Phi}\,{\rm Im}\, \hat\Psi_1]\wedge e^\calf\ .
\eea
Note that  $\xi$ is constant under deformations of $(\Sigma,\calf)$ within the same generalized homology class and thus depends only on the closed string moduli (namely, those contained in the non-integrable pure spinor),
which we always consider frozen in this paper.

It would be interesting to understand whether or not or under what additional conditions the necessary condition
\bea\label{zeroFI}
\xi=0
\eea
is also sufficient to ensure the existence of a solution to the D-flatness condition (\ref{dflatness})
inside each orbit of generalized complex cycles generated by the action of $\calg^{\mathbb{C}}$.
The  condition (\ref{zeroFI}) is for example analogous to the necessary stability condition $\int \omega\wedge F=0$
for a $U(1)$ holomorphic gauge connection that is in this case also sufficient to ensure the existence of a solution to the Hermitian Yang-Mills (HYM) equation $\omega\wedge F=0$
(the D-flatness condition in our context) on a K\"ahler four-manifold with K\"ahler form $\omega$ (see for example \cite{GSW}).
In general, from the results obtained on the same problem in the case of ordinary Calabi-Yau manifolds,
we expect that some stability condition must be added to the condition (\ref{zeroFI}).
The correct notion of stability for these generalized complex cycles should result
from an extension of the results and methods applied to the analogous question for Lagrangian cycles
(see for example \cite{joycelag,thomas1,thomas2}). For example, a natural physical stability condition for a certain
generalized complex cycle $(\Sigma,\calf)$ with vanishing Fayet-Iliopoulos term may be that it cannot split
in two generalized complex cycles with vanishing Fayet-Iliopoulos terms.

If we would know the correct definition of stability, we should only consider the subsector $\calc^{\rm stable}_{\rm hol,\xi=0}\subset \calc_{\rm hol}$ of stable generalized complex cycles with vanishing Fayet-Iliopoulos terms, obtaining that the true moduli space of supersymmetric/calibrated generalized cycles is given by
\bea
\calm=\calc^{\rm stable}_{\rm hol,\xi=0}/\calg^{\mathbb{C}}\ .
\eea

We postpone these non-trivial issues to future investigations.
It is however clear that generalized complex geometry seems to be the correct language
to unify and extend the results already obtained for A- and B-branes on ordinary Calabi-Yaus
(see for example \cite{aspinwallDbranes} for a recent review) to backgrounds with fluxes.

\section{K\"ahler potentials}
\label{k}

This paper is focused  on the study of the flat directions of the moduli space of
supersymmetric/calibrated space-time filling D-branes. We have found that the massless spectrum around
supersymmetric configurations, and in general the full superpotential characterizing the system,
depends only  on the generalized complex structure associated to the $d_H$-closed pure spinor $\hat\Psi_2$.

However,  if one wants to give a complete $\caln=1$ four-dimensional description of the system,
the superpotential provides only partial information, that must be completed by the information encoded
in the K\"ahler potential. The latter depends instead on the {\em non-integrable} generalized almost complex structure
through the associated non-$d_H$-closed pure spinor $\hat\Psi_1$. We have not solved the problem of
identifying the complete moduli space of supersymmetric D-branes, and so we cannot hope to provide a K\"ahler potential
describing the sigma model associated to it. We can however focus on a particular supersymmetric configuration $(\Sigma,\calf)$, and look for the $\caln=1$ theory describing the small fluctuations around it. The appropriate metric entering the kinetic term of general fluctuations around any  generalized cycle $(\Sigma,\calf)$ was given in \cite{branesuppot}, and is recalled in equation (\ref{metric}) of appendix \ref{mm} in the case that $(\Sigma,\calf)$ is supersymmetric. Take now  a base $\mathbb{X}_i$ of  $d_{L_{(\Sigma,\calf)}}$-closed sections of $\caln^{0,1}_{(\Sigma,\calf)}$ which furthermore satisfy the complexified D-flatness condition (\ref{complexcond}) or, in other words, $\mathbb{X}_i$ are  harmonic representatives (with respect to the generalized Laplacian (\ref{laplacian})) of a base for   $H^1(L_{(\Sigma,\calf)})$. Using the metric (\ref{metric}) we can write the following K\"ahler potential  for the  massless four-dimensional chiral fields $\phi^i$ associated to the base $\mathbb{X}_i$:\footnote{Again, we are ignoring the proper dimensionful factors which can easily be introduced, see \cite{branesuppot}.}
\bea\label{kahler}
\calk=\calk_{i\bar k}\phi^i\bar\phi^{\bar k}\ ,
\eea
where
\bea\label{kmetric}
\calk_{i\bar k}=-i\int_{\Sigma}P_\Sigma[e^{2A-\Phi}\mathbb{X}_i\cdot \bar{\mathbb{X}}_{\bar k} \cdot {\rm Im}\,\hat\Psi_1]\wedge e^\calf\ ,
\eea
is the associated K\"ahler metric. Let us stress that the K\"ahler potential (\ref{kahler}) depends on the warp factor, which is in general non-trivial.

As an example, let us now see how the K\"ahler metric looks like for
a D$6$-brane wrapping a special Lagrangian cycle in a IIA background.
Differently from section \ref{examples}, we explicitly need here the non-integrable pure spinor $\hat\Psi_1$.
We have ${\rm Im} \, \hat\Psi_1={\rm Im}\,(e^{i\varphi}\Omega)$, where $\varphi$ is a phase determined by the two internal
six-dimensional spinors generating the supersymmetry. The  base of massless fluctuations $\mathbb{X}_i$ is given by   elements of the form (\ref{agnor}) that must be closed with respect to $d_{L_{(\Sigma,\calf=0)}}$ and $d_{L_{(\Sigma,\calf=0)}}^\dagger$, i.e.
\bea
dP_\Sigma[\iota_{V_i}\omega]=0\quad,\quad d \star_3P_\Sigma[e^{2A-\Phi}\iota_{V_i}\omega]=0\ ,
\eea
where $\star_3$ is the standard Hodge-star operator computed using the induced metric on the three-cycle.
 The K\"ahler metric (\ref{kmetric}) takes the form
\bea
\calk_{i\bar k}=2\int_\Sigma e^{2A-\Phi} P_\Sigma[\iota_{V_i}\omega\wedge \iota_{\bar V_{\bar k}}{\rm Im}\,(e^{i\varphi}\Omega)]=2\int_\Sigma e^{2A-\Phi} P_\Sigma[\iota_{V_i}\omega]\wedge \star_3 P_\Sigma[\iota_{\bar V_{\bar k}}\omega]\ ,
\eea
providing the generalization of the usual metric for special Lagrangian branes to warped flux backgrounds.
For a case by case discussion of other D-branes see \cite{branesuppot}.


\end{appendix}

\newpage


\end{document}